\documentclass[12pt,a4paper]{article}%
\pdfoutput=1
\usepackage{jheppub}
\usepackage{amsmath,amssymb,amsfonts,wasysym}
\usepackage{color}
\usepackage{mathrsfs}
\usepackage{simplewick}
\usepackage[bf,footnotesize]{caption2}
\usepackage{bbm}
\usepackage{ulem}
\usepackage{url}
\usepackage{latexsym,epsfig,amssymb,amsmath,slashed}

\definecolor{rossos}{cmyk}{0,1,1,0.55}
\definecolor{bluscuro}{rgb}{0.15, 0.2, .85}
\definecolor{bluchiaro}{cmyk}{1,.3,0.,0.1}

\newcommand{\bc}{\begin{center}}
\newcommand{\ec}{\end{center}}
\newcommand{\MET}{E\llap{/\kern1.5pt}_T}
\newcommand{\pMET}{{\bf p}\llap{/\kern1.5pt}_T}
\newcommand{\bea}{\begin{eqnarray}}
\newcommand{\eea}{\end{eqnarray}}

\catcode`@=11
\def\marginnote#1{}
\newcommand{\nn}{\nonumber}
\newcount\hour
\newcount\minute
\newtoks\amorpm
\hour=\time\divide\hour by60
\minute=\time{\multiply\hour by60 \global\advance\minute by-\hour}
\edef\standardtime{{\ifnum\hour<12 \global\amorpm={am}%
\else\global\amorpm={pm}\advance\hour by-12 \fi
\ifnum\hour=0 \hour=12 \fi
\number\hour:\ifnum\minute<10 0\fi\number\minute\the\amorpm}}
\edef\militarytime{\number\hour:\ifnum\minute<10 0\fi\number\minute}
\def\draftlabel#1{{\@bsphack\if@filesw {\let\thepage\relax
\xdef\@gtempa{\write\@auxout{\string
\newlabel{#1}{{\@currentlabel}{\thepage}}}}}\@gtempa
\if@nobreak \ifvmode\nobreak\fi\fi\fi\@esphack}
\gdef\@eqnlabel{#1}}
\def\@eqnlabel{}
\def\@vacuum{}
\def\draftmarginnote#1{\marginpar{\raggedright\scriptsize\tt#1}}
\def\draft{\oddsidemargin 0.0truein
\def\@oddfoot{\sl ES, preliminary notes \hfil
\rm\thepage\hfil\sl\today\quad\militarytime}
\let\@evenfoot\@oddfoot \overfullrule 3pt
\let\label=\draftlabel
\let\marginnote=\draftmarginnote
\def\@eqnnum{(\theequation)\rlap{\kern\marginparsep\tt\@eqnlabel}%
\global\let\@eqnlabel\@vacuum} }
\catcode`@=12
\newcommand{\be}{\begin{equation}}
\newcommand{\ee}{\end{equation}}
\newcommand{\beq}{\begin{equation}}
\newcommand{\eeq}{\end{equation}}
\newcommand{\beqs}{\begin{eqnarray}}
\newcommand{\eeqs}{\end{eqnarray}}

\def\({\left(}
\def\){\right)}
\def\<{\langle}
\def\>{\rangle}

\def\be{\begin{equation}}
\def\ee{\end{equation}}
\def\bry{\begin{array}}
\def\ery{\end{array}}
\def\bes{\begin{subequations}}
\def\ees{\end{subequations}}
\def\bit{\begin{itemize}}
\def\eit{\end{itemize}}
\def\ben{\begin{enumerate}}
\def\een{\end{enumerate}}

\definecolor{grey}{rgb}{0.6,0.6,0.6}
\definecolor{fuchsia}{rgb}{1,0,1}

\newcommand{\urdrdr}{(u_R d_R d_R)^2}
\newcommand{\urdrdl}{(u_R d_R d_L)^2}
\newcommand{\uldldr}{(u_L d_L d_R)^2}
\newcommand{\urdrsr}{(u_R d_R s_R)^2}

\newcommand{\onecontr}[6]{
\acontraction{]}{#1}{}{#2}
\bcontraction{[ #1 #2}{#3}{][}{#4}
\acontraction{[#1 #2 #3][#4}{#5}{}{#6}
[ #1 #2 #3][ #4 #5 #6]}

\newcommand{\threecontr}[6]{
\bcontraction {[ #1 #2}{ #3}{][}{ #4}
\acontraction {[}{#1}{ #2 #3][ #4}{ #5}
\acontraction[1.2ex] {[ #1}{ #2}{ #3][ #4 #5}{#6}
[ #1 #2 #3][ #4 #5 #6]}

\newcommand{\threecontrH}[6]{
\bcontraction {[ #1 #2}{ #3}{][}{ #4}
\acontraction {[}{#1}{ #2 #3][ #4}{ #5}
\acontraction[1.7ex] {[ #1}{ #2}{ #3][ #4 #5}{#6}
[ #1 #2 #3][ #4 #5 #6]}

\title{Baryon number violation in supersymmetry: $n\,{-}\,\bar n$ oscillations as a probe beyond the LHC}
\author[a]{Lorenzo Calibbi}
\author[b]{Gabriele Ferretti}
\author[c]{David Milstead}
\author[b,d,e]{Christoffer Petersson}
\author[c]{Ruth P\"ottgen}
\affiliation[a]{State Key Laboratory of Theoretical Physics, Institute of Theoretical Physics,\\Chinese Academy of Sciences,
Beijing 100190, P.~R.~China}
\affiliation[b]{Department of Physics, Chalmers University of Technology,\\ 412 96 G\"oteborg, Sweden}
\affiliation[c]{Fysikum, Stockholms Universitet, SE-106 91 Stockholm}
\affiliation[d]{Physique Th\'eorique et Math\'ematique, Universit\'e Libre de Bruxelles,\\ C.P. 231, 1050 Brussels, Belgium}
\affiliation[e]{International Solvay Institutes, Brussels, Belgium}
\emailAdd{calibbi@itp.ac.cn}
\emailAdd{ferretti@chalmers.se}
\emailAdd{christoffer.petersson@chalmers.se}
\emailAdd{milstead@fysik.su.se}
\emailAdd{ruth.pottgen@fysik.su.se}

\abstract{We study baryon number violation in $R$-parity violating supersymmetry with focus on $\Delta B=2$ processes which allow neutron--anti-neutron ($n\,{-}\,\bar n$) oscillations. We provide prospects for going beyond the present limits by means of a new search for $n\,{-}\,\bar n$  oscillations. The motivation is the recently proposed $n\,{-}\,\bar n$ oscillation experiment at the European Spallation Source in Lund, which is projected to be able to improve the current bound on the transition probability in the quasi-free regime by three orders of magnitude. 
We consider various processes giving rise to baryon number violation and extract the corresponding simplified models, including only the relevant superpartners and couplings. In terms of these models we determine the exclusion limits from LHC searches as well as from searches for flavor transitions, CP violation and di-nucleon decays. We find that, for certain regions of parameter space, the proposed $n\,{-}\,\bar n$ experiment has a reach that goes beyond all other experiments, as it can probe gluino and squark masses in the multi-TeV range.}

\begin{document}
\maketitle
%%%%%%%%%%%%%%%%%%%%%%%%%%%%%%%%%%%%%%%
\setcounter{page}{2}

\section{Introduction}

Baryon number violation (BNV) is needed to explain the observed matter-antimatter asymmetry of the universe~\cite{Sakharov:1967dj}, motivating experimental searches for BNV processes. 
The Standard Model (SM) of particle physics predicts BNV to occur only via rare non-perturbative electroweak processes~\cite{Adler:1969gk,'tHooft:1976fv}. Only the difference between baryon and lepton number, $B-L$, is respected in the SM, whereas $B$ and $L$ are separately broken by non-perturbative effects.
However, within the SM, these effects are exceedingly small, and an experimental observation of a BNV process would imply direct evidence of physics beyond the SM.

Baryon number conservation in the SM at the perturbative level is a consequence of the gauge symmetries and the specific matter content, hence it is a so-called  ``accidental" symmetry. High precision tests of the Equivalence Principle~\cite{Schlamminger:2007ht} have so far excluded a long range force coupled to baryon number and thus a local gauge symmetry protecting baryon number. On the other hand, BNV is a generic feature of a number of theories that extend the SM. In the context of supersymmetry (SUSY), BNV theories are included in the class referred to as $R$-parity violation (RPV)~\cite{Barbier:2005rr,Mohapatra:2015fua}.

Many BNV searches have targeted proton decay. In this context, owing largely to the need to ensure angular momentum conservation, such processes must violate both baryon and lepton number simultaneously. A promising BNV-only observable is the conversion of a neutron to an antineutron: a process that would require a change of two units in the baryon number, $|\Delta B|=2$.  Searches have been made for free neutron oscillations and anomalous nuclear decays, under the neutron oscillation or di-nucleon-decay hypothesis~\cite{Phillips:2014fgb,Mohapatra:2009wp}. The Super-Kamiokande experiment~\cite{Abe:2011ky} has set a limit of $1.9 \times 10^{32}$~years for the oscillation of bound neutrons in $^{16}O$, translating, after some assumptions on the nuclear suppression factor, to an indirect estimate of the free $n-\bar{n}$ oscillation time limit of $2.7 \times 10^{8}$~s. The currently best direct measurement of the free $n-\bar{n}$ oscillation time, done by Institut Laue-Langevin (ILL) in Grenoble, sets a bound at $0.86 \times 10^{8}$~s~\cite{BaldoCeolin:1994jz}.

The experiments at the Large Hadron Collider have also made a number of searches e.g.~for anomalous multijet production, at centre-of-mass energies of 8~TeV and 13~TeV, which are sensitive to BNV processes. Sensitivity is also given by precision measurements of flavour-changing processes in the Kaon and Beauty sectors. A new experiment was recently proposed~\cite{EOInnbar} to search for $n-\bar{n}$ oscillations at the European Spallation Source (ESS) in Lund, Sweden, which could extend the sensitivity to the neutron-antineutron transition probability by up to three orders of magnitude compared to the ILL bound  (see also \cite{Milstead:2015toa}). In this paper we quantify how the various measurements impose constraints on BNV-processes and assess the reach of the proposed experiment.

The work is carried out in the framework of RPV SUSY. RPV models have become an attractive research area in light of the lack of the characteristic SUSY
signatures, involving large amount of missing energy, at the LHC. RPV models evade these constraints by allowing the lightest SUSY particle (LSP) to decay into ordinary SM particles. Particularly interesting for $n-\bar n$ oscillations is the case of baryonic RPV, where only $B$ violating couplings are permitted.
In models of this type, proton decay is perturbatively forbidden and the first baryon number violating processes arise at $\Delta B=2$, namely $n-\bar n$ oscillations and di-nucleon decays \cite{Zwirner:1984is,Barbieri:1985ty}. The presence of RPV couplings also give rise to a plethora of other possible effects, from flavour and CP violation to collider signatures.

The paper is organised as follows: In Section~2 we briefly present the six-quark (dimension nine) $\Delta B=2$ operators contributing to $n-\bar n$ oscillations or di-nucleon decay, arising in RPV models. (A more systematic and model independent overview is found in the Appendix). In Section~3 we present the class of RPV models under consideration and the notation used throughout the paper. In Section~4 we present the bounds on such theories arising from flavour physics and CP violation, di-nucleon decay and LHC searches. Section~5 contains the study of $n-\bar n$ oscillation in this context and the comparison with the previous searches.  We show that the proposed experiment at ESS can significantly extend the reach of such searches and test regions of parameter space  otherwise inaccessible. In Section~6 we discuss additional possible contributions to $n-\bar n$ oscillations arising from non-renormalizable operators and in Section~7 we conclude. 

\section{Operators contributing to $n-\bar{n}$ oscillation}
\label{operators}

The operators of interest for $n-\bar{n}$ oscillations and di-nucleon decay in the RPV context are the following:
\beqs
      \urdrdr & \equiv & \epsilon_{abc}  u_{R\dot\alpha}^a d_R^{\dot\alpha b} d_{R\dot\gamma}^c \;  \epsilon_{def} u_{R\dot\beta}^d d_R^{\dot\beta e} d_R^{\dot\gamma f} \nn\\
          \urdrdl & \equiv & \epsilon_{abc}  u_{R\dot\alpha}^a d_R^{\dot\alpha b} d_L^{\gamma c}\;  \epsilon_{def} u_{R\dot\beta}^d d_R^{\dot\beta e} d_{L \gamma}^f \nn\\
          \uldldr & \equiv & \epsilon_{abc}  u_{L}^{\alpha a} d_{L\alpha}^{ b} d_{R \dot\gamma}^c\;  \epsilon_{def} u_{L}^{\beta d} d_{L\beta}^{e} d_{R}^{\dot\gamma f} \nn\\
      \urdrsr & \equiv & \epsilon_{abc}  u_{R\dot\alpha}^a d_R^{\dot\alpha b} s_{R\dot\gamma}^c \;  \epsilon_{def} u_{R\dot\beta}^d d_R^{\dot\beta e} s_R^{\dot\gamma f}. \label{relevantops}
\eeqs
We use two component notation throughout the paper. $a,b,\dots$ are colour indices, $\alpha, \beta,\dots$ left-handed (LH) Weyl indices and
$\dot\alpha, \dot \beta,\dots$ right-handed (RH) ones. The second and third operator are Parity conjugate of each other. The last operator contributes only to di-nucleon decay $NN\to KK$ while the first three contribute to both $n-\bar{n}$ oscillation and di-nucleon decay $NN\to \pi\pi$. (The process $NN\to K\pi$ is never of interest for the models we consider.)

These are just a small set of all the independent $\Delta B=2$ operators that can be constructed and we review their classification in Appendix~A. For now it suffices to note that their renormalization has been computed to leading~\cite{Caswell:1982qs} and subleading~\cite{Buchoff:2015qwa} order. To leading order  in $\alpha_s$ the operators in~\eqref{relevantops} do not mix, the second and third operator are not renormalized at all, while the first and the last are suppressed by about $60\%$ in going from a BSM scale,  if taken to be 10\,TeV, 
down to the nucleon mass scale.

In application to $n{-}\bar{n}$ oscillations, denoting by $\mathcal O$ any dimension nine operator mediating the oscillation, e.g.~one of the first three operators in (\ref{relevantops}), one is interested in the Hamiltonian matrix element ``$\langle n|\mathcal{O}|\bar n\rangle$'' between the $n$ and $\bar n$ defined via
\beq
     \langle n, \mathbf{p} |\int \mathrm{d}^3\mathbf{r}\,\mathcal{O}(\mathbf{r}, t=0)|\bar n,
     \mathbf{q}\rangle = \langle n|\mathcal{O}|\bar n\rangle (2\pi)^3 2 E \delta^{(3)}(\mathbf{p}-\mathbf{q})
\eeq
taking the zero momentum limit.

In applications to di-nucleon decay to e.g.~Kaons, one considers instead the S-matrix element ``$\langle N N'|\mathcal{O}|K K'\rangle$'' between two nucleons and two Kaons defined via
\beq
     \langle N, \mathbf{p}; N', \mathbf{p}' |\int \mathrm{d}^4 x\,\mathcal{O}(x)| K, \mathbf{q}; K', \mathbf{q}'\rangle = i \langle N N'|\mathcal{O}|K K'\rangle (2\pi)^4 \delta^{(4)}(p + p' - q - q')
\eeq
taking the zero momentum limit of the nucleons. In this case $\mathcal O$ is any dimension nine operator mediating the $NN\to KK$ transition, e.g.~the last operator in  (\ref{relevantops}).
With the relativistic normalization $\langle \mathbf{p} | \mathbf{q}\rangle = (2\pi)^3 2 E \delta^{(3)}(\mathbf{p}-\mathbf{q})$ for the single particle states, it can be seen that, dimensionally,
$\langle n|\mathcal{O}|\bar n\rangle = C \Lambda_{\mathrm{QCD}}^6$ and $\langle N N'|\mathcal{O}|K K'\rangle = C'  \Lambda_{\mathrm{QCD}}^5$ for some dimensionless coefficients $C$ and $C'$ depending on the operators and on the process at hand.

\section{Baryon number violating supersymmetry}

In this paper we will consider only RPV SUSY models where baryon number is violated (BRPV) but where lepton number is preserved. 
In such models, proton decay poses no problem, and dark matter could be accommodated by e.g.~axions.  At the renormalizable level, the only additional interaction we can write down, beyond the usual MSSM superpotential, is
\begin{eqnarray}
\label{uddsup}
W_{BRPV}
&=& %\frac{1}{2}
\lambda^{''}_{ijk}\epsilon_{abc}
\bar{U}^a_i \bar{D}^b_j \bar{D}^c_k
\end{eqnarray}
where $i,j,k$ and $a,b,c$ are flavour and colour indices, respectively, and where the dimensionless coupling is antisymmetric in the last two indices, $\lambda^{''}_{ijk}=-\lambda^{''}_{ikj}$.\footnote{Due to the antisymmetry of $\lambda^{''}_{ijk}$, it is common to define the interaction in Eq.~\eqref{uddsup} with a factor of $1/2$ in front. However, in order to compare to bounds previously obtained in the literature, in which the factor of $1/2$ was omitted, we have chosen this normalization.}
This antisymmetry implies that there are 9 independent $\lambda^{''}_{ijk}$-couplings: $\lambda^{''}_{uds}, \lambda^{''}_{udb} \dots $. We will use this explicit notation in terms of the quark/squark flavour when discussing explicit processes. The relevant couplings that can be probed at the $n-\bar n$ experiment under various assumptions are $\lambda^{''}_{uds}$,  $\lambda^{''}_{udb}$  and $\lambda^{''}_{tdb}$.
The superpotential \eqref{uddsup} carries baryon number $-1$, so the couplings   $\lambda^{''}_{ijk}$ violate baryon number by one unit and to obtain $n{-}\bar{n}$ oscillations we need to use the coupling in \eqref{uddsup} twice.

The scalar components of $\bar{U}, \bar{D}, Q$ are denoted by $\tilde{u}^\ast_R , \tilde{d}^\ast_R , (\tilde{u}_L , \tilde{d}_L)$ and the fermion components by $u_R^\dagger, d_R^\dagger, (u_L,d_L)$, which are Weyl fermions that are all left-handed (with respect to the Lorentz group). The superpotential \eqref{uddsup} gives rise to the following component interactions that are relevant for us,
\begin{eqnarray}
\label{uddcomp}
\mathcal{L}_{BRPV} & = & -%\frac{1}{2}
\lambda^{''}_{ijk}\epsilon_{abc} \left(
\tilde{u}^{a}_{R i} d^{b}_{R\, j} d^{c}_{R\, k}+
 u^{a}_{R\, i} \tilde{d}^{b}_{R j} d^{c}_{R\, k}+
 u^{a}_{R\, i} d^{b}_{R\, j} \tilde{d}^{c}_{R k}
\right) +\mathrm{h.c.}
\end{eqnarray}

\begin{figure}[t]
%\begin{center}
\hspace{-2cm} 
%\vspace{1cm}
\includegraphics[width=0.59\textwidth]{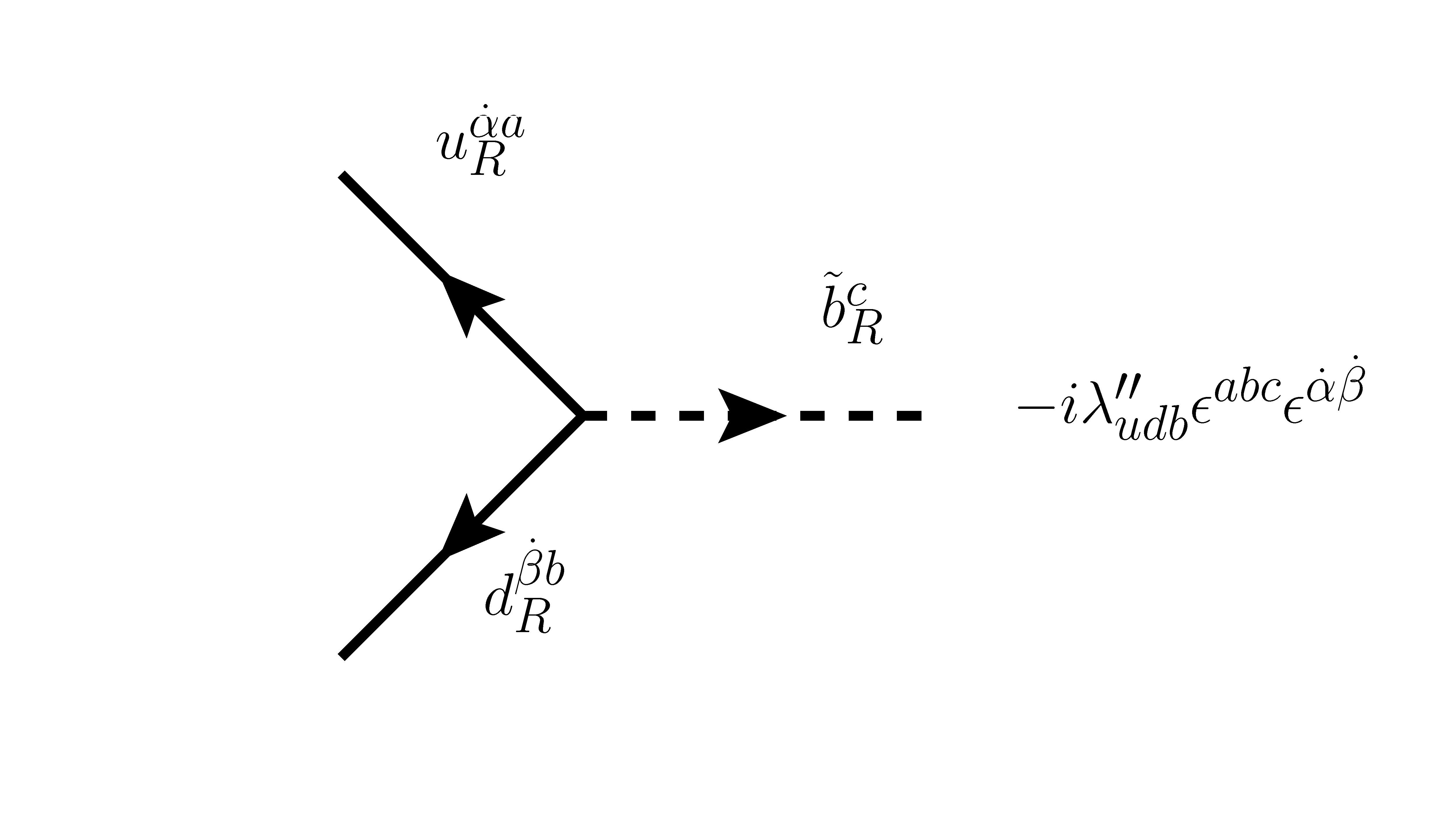}
%\hspace{0.5cm}
%\vspace{-2cm}
\includegraphics[width=0.55\textwidth]{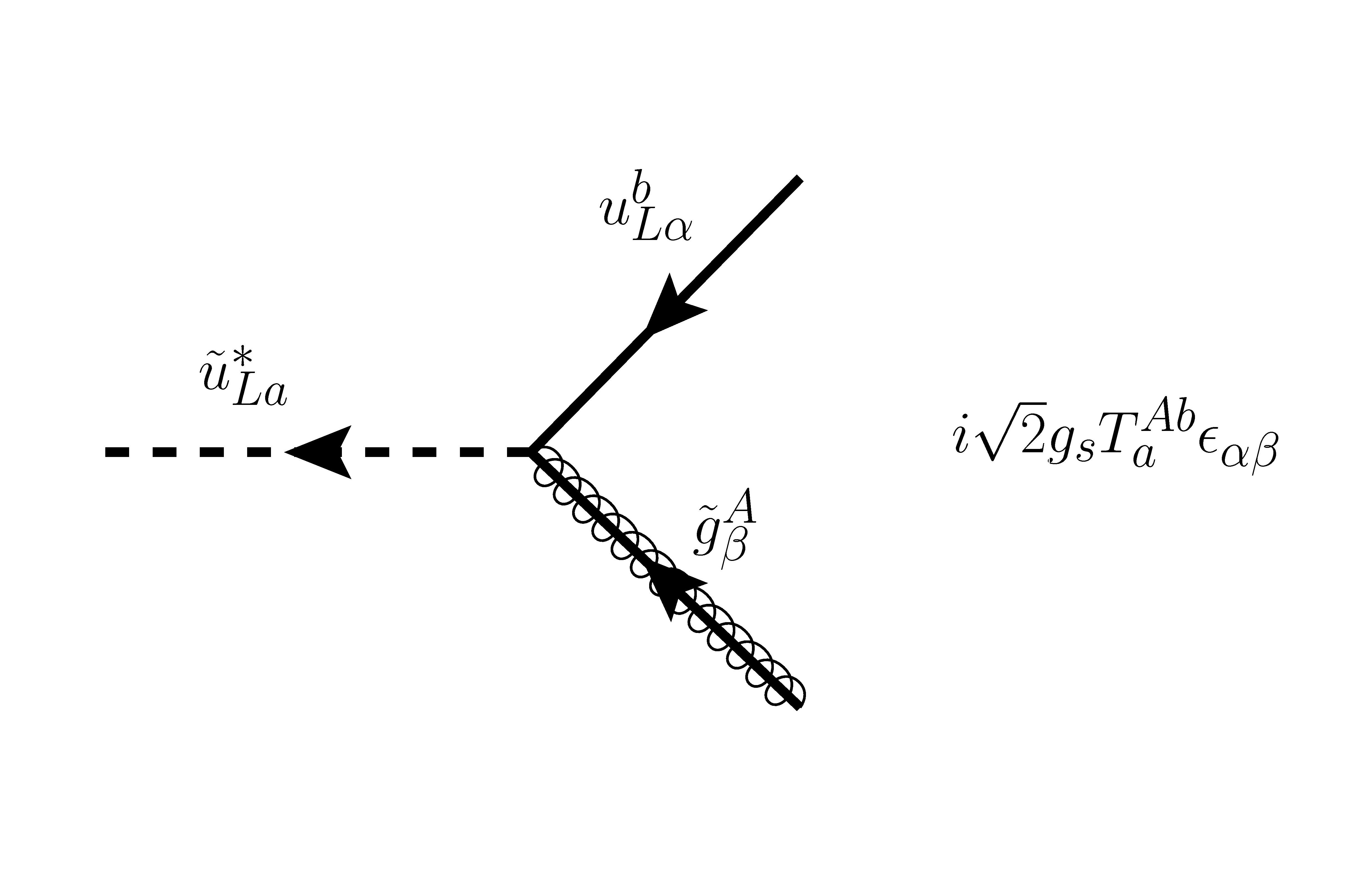}
\caption{Left: Example of an RPV vertex. Right: Example of a quark-squark-gluino vertex. \label{fig:vertices}}
%\end{center}
\end{figure}

When writing the diagrams corresponding to the various processes, we will follow the convention that arrows on fermionic lines represent chirality: LH (undotted)  indices correspond to a line entering a vertex and vice versa for RH (dotted) ones. Scalar lines are also oriented according to the holomorphy of the corresponding fields in a way that Yukawa vertices from a superpotential have always either three incoming or three outgoing lines. A vertex with a gaugino, on the other hand, has the orientation of the scalar line reversed compared to the two fermionic ones. Examples of vertices following such conventions are shown in  Fig.~\ref{fig:vertices}.

With these conventions, a mixing term between two squarks of the same handness, such as e.g. $\tilde b_R^* d_R$ will preserve the orientation of the arrow on the scalar lines while a term switching handness, such as  $\tilde b_R^* b_L$ will reverse it. Fermion masses are always orientation reversing, of course.

 \section{Constraints}
 \subsection{Flavour and CP violation}
\label{sec:FCNC}
Because of the antisymmetric structure of the $\lambda^{''}_{ijk}$ couplings, non-vanishing RPV interactions of first generation quarks must involve second or third generation squarks, $\tilde{s}_R$ or $\tilde{b}_R$.
As we are going to see in the next section, this implies that $n-\bar{n}$ oscillations will arise only in presence of mixing among different squark flavours.
Flavour violation in the squark sector is tightly constrained by meson oscillations and other flavour-changing neutral current (FCNC)
and possibly CP-violating (CPV) processes, see e.g.~\cite{Altmannshofer:2009ne}.
Here, we are going to discuss the constraints that can affect the predictions for $n-\bar{n}$ oscillations in RPV models, presenting the bounds
in the terms of customary mass-insertion parameters:
\begin{equation}
\label{eq:deltas}
\left( \delta^d_{RR}\right)_{ij} \equiv  \frac{(\tilde{m}^2_{D})_{ij}}{\overline{m}^2_{D}},\quad
\left( \delta^d_{LL}\right)_{ij} \equiv  \frac{(\tilde{m}^2_{Q})_{ij}}{\overline{m}^2_{Q}},\quad
\left(\delta^d_{LR}\right)_{ij} \equiv \frac{m_{j} A^d_{ij}}{\overline{m}_{D}\overline{m}_{Q}},
\end{equation}
where $i\neq j$ and $m_{i}$ are down quark masses; $A^d_{ij}$, $(\tilde{m}^2_{D})_{ij}$, and $(\tilde{m}^2_{Q})_{ij}$ are off-diagonal entries of
the A-term matrix, and the squark mass matrices (RH and LH respectively), expressed in the flavour basis where the down-quark mass matrix is diagonal.
Finally, $\overline{m}_{D}$ and $\overline{m}_{Q}$ are average RH and LH down-squark masses.
These parameters control the degree of mixing among squarks of different generations and can be employed to write the amplitudes of FCNC processes in the so-called mass-insertion approximation (MIA), c.f.~\cite{Gabbiani:1996hi}, which gives accurate results as far as the squarks are almost mass-degenerate and the above parameters are $\ll1$.
 \begin{figure}[t]
\begin{center}
\includegraphics[width=0.4\textwidth]{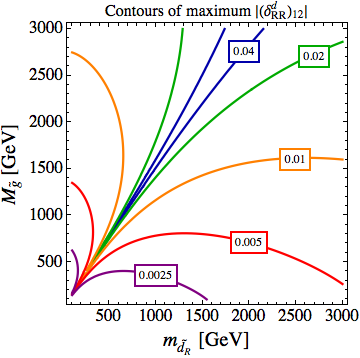}
\hspace{0.4cm}
\includegraphics[width=0.4\textwidth]{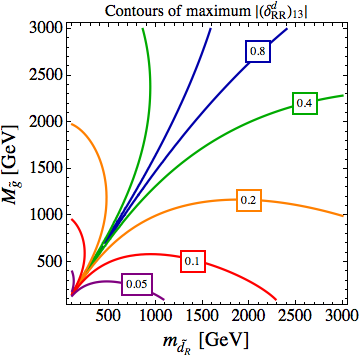}
\includegraphics[width=0.4\textwidth]{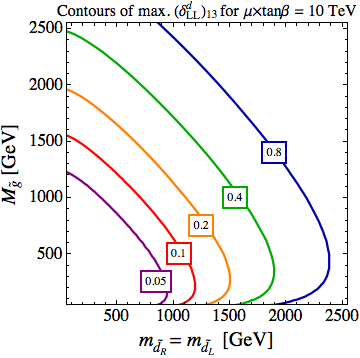}
\hspace{0.4cm}
\includegraphics[width=0.4\textwidth]{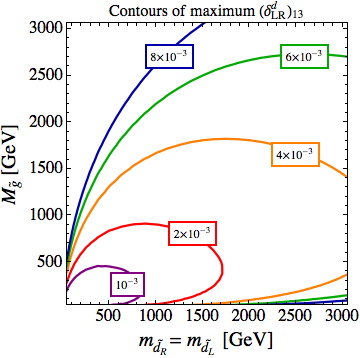}
\caption{Bounds from flavour and CP violation. See the text for details. \label{fig:delta12bound}}
\end{center}
\end{figure}

If flavour violation occurs in the 1-2 sector, this gives rise to contributions to $K-\bar{K}$ mixing that are stringently constrained by the observed Kaon mass splitting $\Delta m_K$ and CP violation parameter $\epsilon_K$. In the upper-left panel of Fig.~\ref{fig:delta12bound}, we show in the plane
($m_{\tilde{d}_R}=m_{\tilde{s}_R},~M_{\tilde g}$) the bound on $\left|\left(\delta^d_{RR}\right)_{12}\right|$ from $\epsilon_K$ obtained assuming an $\mathcal{O}(1)$ CPV phase, i.e.~arg$\left(\left( \delta^d_{RR}\right)_{12}\right)=1$. The bounds have been computed using the expression of the Wilson coefficient of the FCNC operator $(\bar{s}\gamma_\mu P_R d)(\bar{s}\gamma^\mu P_R d)$ given in \cite{Altmannshofer:2009ne} and comparing
the results with the bounds reported in \cite{Isidori:2010kg,Calibbi:2012yj,Calibbi:2012at}.
Similarly, in the case of flavour violation in the 1-3 sector, constraints on $\left|\left(\delta^d_{RR}\right)_{13}\right|$ come from $B-\bar{B}$ mixing:
these are much milder than the analogous ones of the 1-2 sector, as shown in the upper-right panel of Fig.~\ref{fig:delta12bound}. The bounds have
been computed as in the previous case. Values of $\left|\left(\delta^d_{RR}\right)_{13}\right|=\mathcal{O}(1)$, for which the MIA breaks down, are
also displayed: these should be just regarded as indicative of regions of the parameter space where no bound from FCNC processes can be set.

As we are going to see, a class of contributions to $n-\bar{n}$ involves gluinos and down squarks of both RH and LH kinds,
featuring a LR squark chirality flip and flavour violation in the LH sector,
or both the LR and the flavour mixing directly given at the same time by flavour-violating A-terms, i.e.~$\left(\delta^d_{LR}\right)_{ij}$.
Bounds on $\left|\left(\delta^d_{LL}\right)_{13}\right|$ from $B-\bar{B}$ mixing are similar to those  for $\left|\left(\delta^d_{RR}\right)_{13}\right|$
shown in the upper-right panel of Fig.~\ref{fig:delta12bound}.
A more stringent flavour constraint on this scenario comes from $b\to d\gamma$ transitions,
due to sizeable contributions to flavour violating dipole operators induced by the large LR mixing $\propto \mu\times\tan\beta$.
The corresponding bound for the illustrative case of $\mu\times\tan\beta=10$ TeV is shown in the lower-left panel of Fig.~\ref{fig:delta12bound}.
The Wilson coefficient of the dipole operators have been computed in the MIA as in \cite{Altmannshofer:2009ne}, and
employed to obtain the BR($b\to d\gamma$) using the expressions of \cite{Hurth:2003dk}.
The resulting bound on $\left|\left(\delta^d_{LL}\right)_{13}\right|$ has been obtained as in \cite{Crivellin:2011ba}.
Similarly, $b\to d\gamma$ strongly constrains $\left(\delta^d_{LR}\right)_{13}$, as shown in the lower-right panel of Fig.~\ref{fig:delta12bound}.

 \subsection{Di-nucleon decays}
 \label{sec:NN}
\begin{figure}[t]
%\begin{center}
\hspace{-0.6cm}
\includegraphics[width=0.52\textwidth]{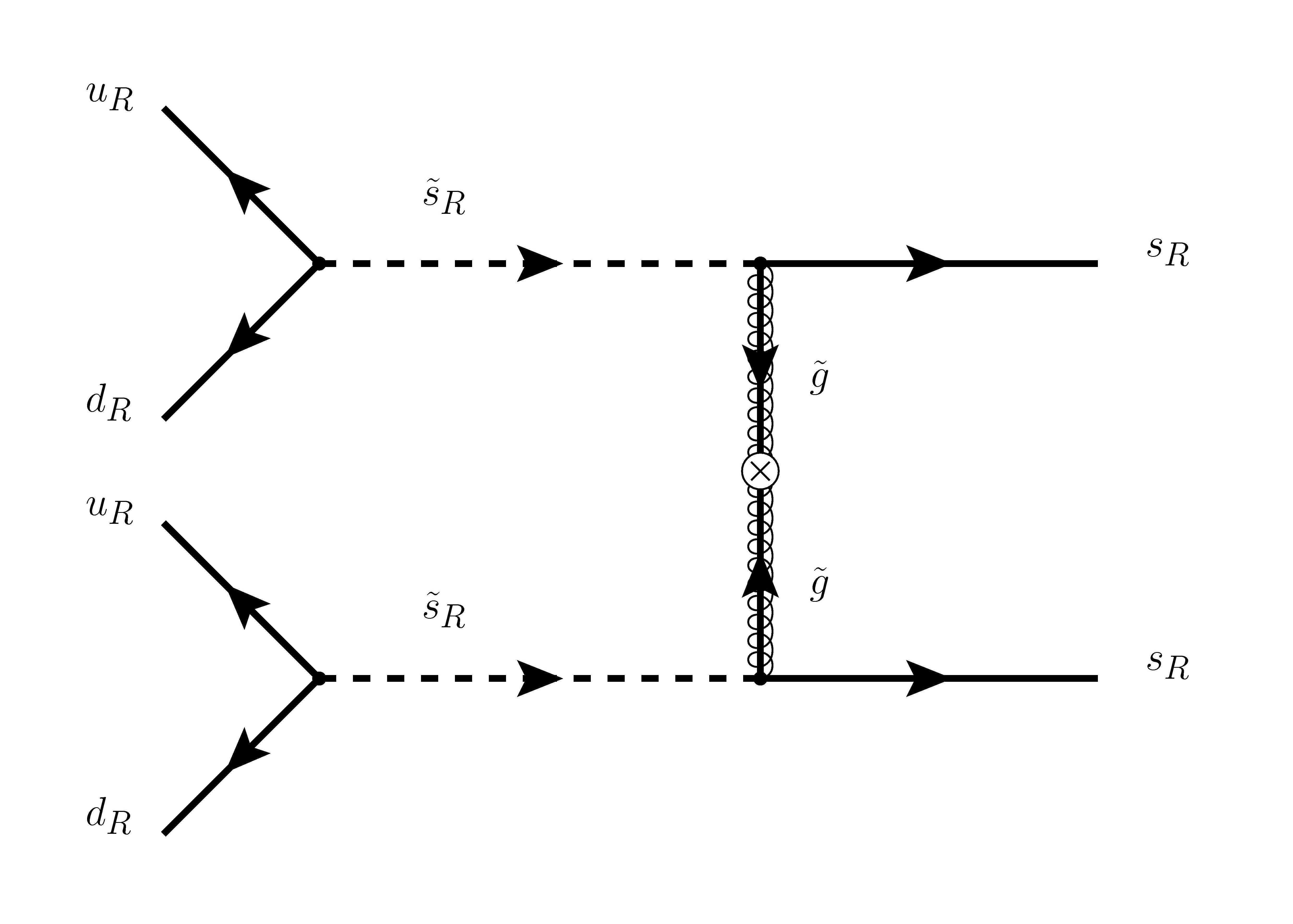}%\hspace{-0.2cm}
\includegraphics[width=0.52\textwidth]{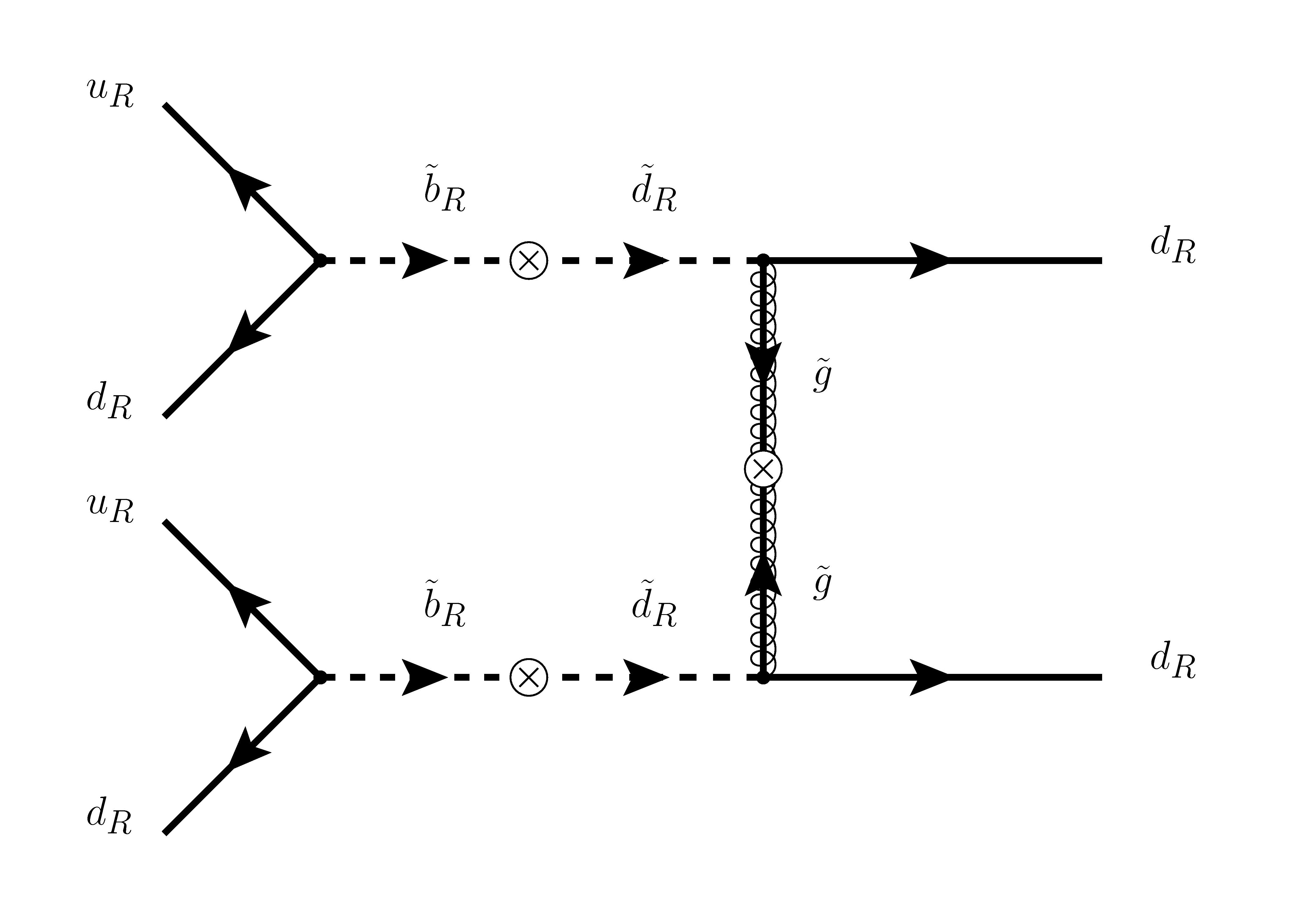}
\caption{Di-nucleon decay diagrams for the processes $NN{\to}KK$ (left) and $nn\to \pi^0\pi^0$ (right). \label{fig:NNKK}}
%\end{center}
\end{figure}
A stringent constraint on $\lambda^{''}_{uds}$ comes from the double nucleon decay to two Kaons, $NN{\to}KK$ \cite{Barbieri:1985ty,Goity:1994dq}. The corresponding diagram is shown in Fig.~\ref{fig:NNKK} (left).
This process violates both baryon and strangeness number by two units and arise from the following dimension 9 operator:
\begin{eqnarray}
\label{NNKK}
\mathcal{L}_{NN{\to}KK} & = & \frac{4}{3} \frac{g_s^2 (\lambda_{uds}^{''})^2}{M_{\tilde{g}} \,m_{\tilde{s}_R}^4}
\urdrsr
+\mathrm{h.c.},
\end{eqnarray}
where  $\urdrsr$ is given in (\ref{relevantops}), $g_s$ is the strong coupling, $M_{\tilde{g}}$  and $m_{\tilde{s}_R}$ are the masses of the SUSY particles involved in the process
(cf.~Fig.~\ref{fig:NNKK}, left): the gluino and the RH strange squark, respectively.
The expression for the nuclear matter lifetime reads \cite{Goity:1994dq}:
\begin{equation}
\label{tauNNKK}
\tau_{NN{\to}KK}=  \frac{m_N^2\,M_{\tilde{g}}^2 \,m_{\tilde{s}_R}^8}{8\,\pi \,\alpha_s^2 \,(\lambda_{uds}^{''})^4 \,\rho_N \,
\langle KK|\urdrsr| NN\rangle^2}
\end{equation}
where $m_N$ is the nucleon mass, $\rho_N$ the nuclear matter density,
$\rho_N=0.25$\,fm${}^{-3}$, and $\alpha_s\equiv g_s^2/4\pi$.

\begin{figure}[t]
\begin{center}
\includegraphics[width=0.4\textwidth]{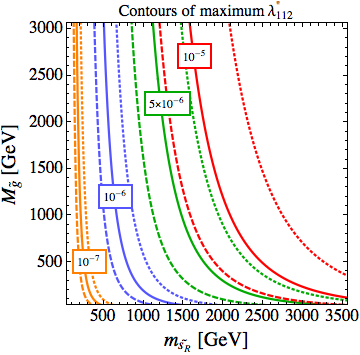}
\hspace{0.5cm}
\includegraphics[width=0.4\textwidth]{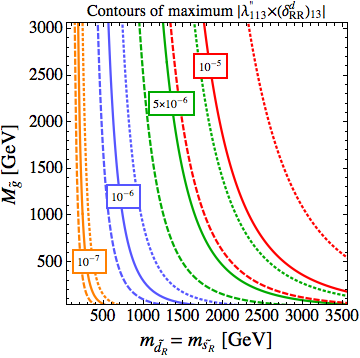}
\caption{Bounds from di-nucleon decays on RPV couplings from limits on $NN\to KK$ (left) and $NN\to \pi\pi$ (right). See the text for details. \label{fig:NNKK2}}
\end{center}
\end{figure}

The most recent limit can be extracted from a search performed by Super-Kamiokande for the decay ${}^{16}O{\to} {}^{14}C\,K^+K^+$  \cite{Litos:2014fxa},
corresponding to the mode $pp\to K^+ K^+$.
The resulting limit on the di-nucleon lifetime is $1.7{\times}10^{32}$ years.
In the left panel of Fig.~\ref{fig:NNKK2} we see contours of the resulting bound on $\lambda_{uds}^{''}$ displayed
on strange squark-gluino mass plane.
Solid lines correspond to the following choice for the hadronic matrix element: $\langle NN|\urdrsr|KK\rangle \equiv (150\mbox{ MeV})^5$. In order to show the large uncertainty due to this poorly known quantity, we also display as dashed (dotted) lines
the bounds obtained dividing (multiplying) the value of $\langle NN|\urdrsr|KK\rangle$ by a factor of 3, thus the area between dashed and dotted lines correspond to an order of magnitude variation of the matrix element.

Super-Kamiokande recently set limits on di-nucleon decays to pions, among which the most stringent is
$\tau_{nn\to \pi^0\pi^0} >4.04{\times}10^{32}$ years \cite{Gustafson:2015qyo}. This constraint is relevant  for $\lambda_{uds}^{''}$  provided that the strange
squark $\tilde{s}_R$ mixes with $\tilde{d}_R$. This is indeed a necessary condition to give rise to $n-\bar{n}$ oscillations, as we will see in the next section.
However,  $nn\to \pi^0\pi^0$ would then constrain the product $\lambda_{uds}^{''}\times (\delta^d_{\rm RR})_{12}$. Hence, given the stringent bounds on
$(\delta^d_{\rm RR})_{12}$ from $K-\bar{K}$ oscillations that we discussed above in section \ref{sec:FCNC} and the fact that the limits on
$nn\to \pi^0\pi^0$ and $pp\to K^+ K^+$ are of the same order of magnitude, $nn\to \pi^0\pi^0$  can not set a more stringent constraint on
$\lambda_{uds}^{''}$  than the direct di-nucleon decays to Kaons that do not require squark flavour mixing.

Instead, $nn\to \pi^0\pi^0$ does give a relevant constraint on $\lambda_{udb}^{''}$
(or rather on $\lambda_{udb}^{''}\times (\delta^d_{\rm RR})_{13}$), which is otherwise unconstrained by di-nucleon decays given that decays to $B$ mesons
are kinematically forbidden. The diagram is shown in Fig.~\ref{fig:NNKK} (right).
The lifetime reads in this case:
\begin{equation}
\label{tauNNpipi}
\tau_{NN{\to}\pi\pi}=  \frac{m_N^2\,M_{\tilde{g}}^2 \,m_{\tilde{d}_R}^4 m_{\tilde{b}_R}^4}{8\,\pi \,\alpha_s^2 \,|\lambda_{udb}^{''}\times(\delta^d_{\rm RR})_{13}|^4\,\rho_N \,
\langle \pi\pi|\urdrdr| NN\rangle^2},
\end{equation}
where $\urdrdr$ is given in (\ref{relevantops}).
The resulting bounds on $\lambda_{udb}^{''}\times (\delta^d_{\rm RR})_{13}$ are displayed in the right panel of Fig.~\ref{fig:NNKK2}, for the same
choices of the hadronic matrix element as in the right panel (cf.~the above discussion).

In general, any theory giving rise to $n-\bar{n}$ oscillation is also inducing $NN\to\pi\pi$, as the same operators contribute to both processes, cf.~Eq.~(\ref{relevantops}). Then, in presence of a Lagrangian term $\mathcal{C\cdot O}$, with $\mathcal{O}$ being one  of those operators, we simply have:
\begin{equation}
\tau_{NN{\to}\pi\pi}=  \frac{32\pi}{9}  \frac{m_N^2}{\rho_N \mathcal{C}^2 \langle \pi\pi| \mathcal{O}| NN\rangle^2}.
\label{eq:din-gen}
\end{equation}
Eq.~(\ref{tauNNpipi}) is a specific example of the above contribution. As we are going to see in the next section, the bounds obtained from this contribution to $NN\to\pi\pi$ tend to be subdominat with respect to those from $n-\bar{n}$. However, both processes are affected by large hadronic uncertainties.

 \subsection{LHC searches}
 \label{sec:LHC}
\begin{figure}[t]
\begin{center}
\includegraphics[width=0.4\textwidth]{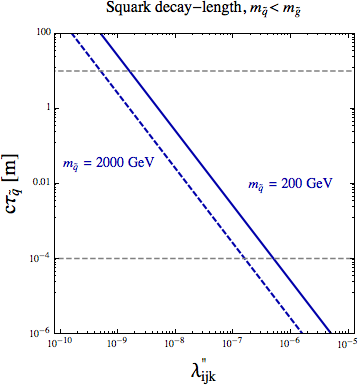}
\hspace{0.5cm}
\includegraphics[width=0.4\textwidth]{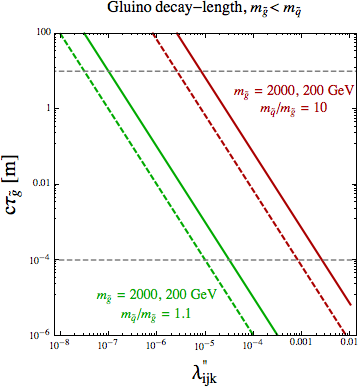}
\caption{The decay lengths of a squark (left) and a gluino (right).\label{fig:decay-lenghts}}
\end{center}
\end{figure}

In the model considered here, the squarks and gluinos can become long-lived due to
weak couplings to SM particles. In the case where the lightest superpartner is a squark, it will necessarily decay into two quarks via an RPV interaction.
The decay width for this process is
\begin{eqnarray}
\label{2-body}
\Gamma (\tilde{q} \to qq) & = & \frac{(\lambda^{''})^2}{8\pi} m_{\tilde{q}}
\end{eqnarray}
where $\lambda^{''}$ is the appropriate RPV coupling and $m_{\tilde{q}}$ the squark mass. The decay length for this case is plotted in Fig.~\ref{fig:decay-lenghts} (left).

In the case where the gluino is lighter than the squarks, the gluino will decay via a 3-body decay, via an off-shell squark, to three quarks with the width
\begin{eqnarray}
\label{3-body}
\Gamma (\tilde{g}\to qqq)& = & \frac{\alpha_s(\lambda^{''})^2 }{256 \pi^3} \frac{m_{\tilde{g}}^5}{m_{\tilde{q}}^4}\,.
\end{eqnarray}
 The corresponding decay length is plotted in Fig.~\ref{fig:decay-lenghts} (right).

In the case where either squarks or gluinos are long lived, they form so-called $R$-hadrons~\cite{Fairbairn:2006gg}. A $R$-hadron
consists of a heavy sparticle and a light quark system. A $R$-hadron with a large lifetime ($c\tau \sim 10$m) would typically propagate through a LHC detector without decaying. It could, however, interact both electromagnetically and strongly with material in the detector. The electromagnetic interactions are well understood and measurements of continuous ionisation energy loss can be used as a search discriminant~\cite{Fairbairn:2006gg}. There are, however, large uncertainties on hadronic scattering processes which can affect the efficiency of a search. For example, a $R$-hadron leaving a charged particle track in an inner detector system can become neutral after charge exchange processes with detector matter and thus pass through an outer muon chamber as a neutral and undetected object~\cite{Kraan:2004tz,deBoer:2007ii,Mackeprang:2009ad}. Such possible processes are studied by the experiments~\cite{Khachatryan:2011ts,Aad:2011yf,Chatrchyan:2012sp,Aad:2012pra}.

 In the conservative approach adopted here, limits on squark and gluino production which are used correspond to hadronic scattering scenarios which provided the smallest efficiency. For lower $c\tau$ values, the $R$-hadrons can decay in the detector and leave a signature of a displaced vertex and decay products emerging from that vertex. For the couplings considered here, a squark (gluino) $R$-hadron would decay to a di-jet (three-jet) system. Searches for non-decaying and decaying long-lived particles were made by the CMS experiment during Run 1, the results of which were converted into excluded regions of lifetime and mass for stops and gluinos in~\cite{Liu:2015bma,Csaki:2015uza} (see also \cite{Cui:2014twa}).  Using these results, exclusion limits on coupling, mixing parameter and sparticle mass were quantified for the models considered in this work.  In addition, CMS results recently obtained at a centre-of-mass energy of 13~TeV~\cite{CMS:2015kdx} were also taken into account to show the impact of the extension in mass exclusions for $R$-hadrons with long lifetimes $c\tau > 10^2$m.

For sufficiently large coupling values, the decays of squarks and gluinos will be prompt and result in a large number of quarks in the final state.
If the gluino is heavier than the (degenerate) squarks, it will decay into a quark and a squark which in turn will decay into two quarks. Thus, for $\tilde{g}\tilde{g}$ production, for example, there will be 6 quarks produced in the decay.
At the LHC experiments, such events will be characterised by a  large number of jets.  \\
In order to extract bounds in the $(m_{\tilde{g}}-m_{\tilde{q}})$-plane from LHC results, a simulation for a simplified RPV SUSY model was done.
This simulation uses \texttt{MadGraph5\_aMC@NLO} \cite{Alwall:2014hca} (version 2.3.3) and \texttt{Delphes} \cite{deFavereau:2013fsa} (version 3.3.0) together with \texttt{PYTHIA8.212} \cite{Sjostrand:2014zea}.  
For the detector simulation, the default \texttt{Delphes} ATLAS card is used, with the only change being that the jet radius parameter is set to 0.4 instead of 0.6.\\
The set of simplified models considered in this work is described in more detail in Sec.~\ref{sec:SUSYinNnbar}.
The different models feature slightly different sparticle contents (cf. Tab.~\ref{tab:modellist}) but this does not change the kinematics and hence the acceptances in the detectors of the LHC experiments.
This has been verified explicitly for the first two models in Tab.~\ref{tab:modellist} by running two separate simulations considering only the respective sparticles (in particular setting all other squark masses to 3\,TeV) and couplings.
The value of the coupling ($\lambda^{''}_{uds}$ in one case and $\lambda^{''}_{udb}$ in the other) was set to $10^{-3}$.
All other couplings are set to zero.
No significant difference in the relevant kinematic distributions was observed.
Therefore, only simulation samples involving the sparticles of model Z$_2$ are used in the following.\\
The squark and gluino masses are scanned over a range from 200\,GeV to 1.4\,TeV and 300\,GeV and 1.5\,TeV, respectively.
A slightly different sensitivity to the different models will result from the difference in the production cross sections.
Samples are generated separately for $\tilde{g}\tilde{g}$- , $\tilde{g}\tilde{q}$- , $\tilde{q}\tilde{q}$- and $\tilde{q}\bar{\tilde{q}}$-production.
The cross section for each process (both with and without the sbottom) is calculated using \texttt{Prospino\,2.1} \cite{Beenakker:1996ch}.

The first LHC measurement that is considered in the case of prompt decays is a search for SUSY particles in final states with a large number of jets, which was conducted by the ATLAS collaboration on 20.3\,fb$^{-1}$ data collected at a centre-of-mass energy of 8\,TeV \cite{Aad:2015lea}.
For this search, different signal regions are defined by requiring at least 7 jets of high transverse momentum and applying different requirements on the number of b-tagged jets.
Model-independent limits on the visible cross section are provided for each of the regions.
The present study considers a signal region which requires each jet to have a transverse momentum above 120\,GeV, but has no additional requirement on the b-tag multiplicity. \\
The same selection is applied to each of the samples to obtain the acceptance.
These acceptances are then multiplied by the production cross section for the respective process, yielding the {\it{visible}} cross section.
The visible cross sections for all four processes are added and the result can be compared to the ATLAS limit, which is 1.9\,fb for this signal region.
Mass points which yield a visible cross section larger than this limit are excluded.

The above analysis is aimed at signals which result in high jet multiplicities, i.e. it is mostly sensitive to $\tilde{g}\tilde{g}$- and $\tilde{g}\tilde{q}$-production and only to a lesser extent to $\tilde{q}\tilde{q}$- and $\tilde{q}\bar{\tilde{q}}$-production.
Limits on the squark mass can be obtained from a CMS search using di-jet pairs in the final state \cite{Khachatryan:2014lpa}.
As mentioned above, in the models considered for this work, the specific squark flavour does not affect the kinematic distributions but only the cross section.
Thus, even though the CMS limits are obtained for models of $\tilde{t}\bar{\tilde{t}}$ production, they are applicable to the models studied here.
Therefore, there was no need to run the event selection on the signal samples, but the CMS limits could be used directly, scaled by the appropriate cross section.

The LHC limits presented here were made with Run 1 and early Run 2 data.
To quantify projected  limits for the large luminosity dataset ($\sim 300$fb$^{-1}$) that ATLAS and CMS are expected to receive by around 2021,  when the proposed ESS experiment would start, is beyond the scope of this paper. However, it can be conservatively estimated that limits on squark and gluino masses would increase by up to ~1000~GeV, as has been estimated by the LHC experiments for a range of SUSY searches~\cite{CMS:2013xfa,ATLASproj:2013}. Furthermore, some of the searches considered in this paper (long-lived particles and displaced jets) require detector signals which are received later than those which would be expected from particles produced at the primary interaction point and which move at around light speed. This can present a special challenge for triggering and read-out as late signals can be associated to the wrong bunch crossing and lost. As the long-lived sparticle masses increase (and the average speed is thus reduced) such losses can become more severe. It would therefore not be expected that these searches would achieve a greater gain in sensitivity than the searches for prompt SUSY signals.

 \section{Contributions to $n{-}\bar{n}$ oscillations from supersymmetry}
\label{sec:SUSYinNnbar}
\begin{table}
\begin{center}
\renewcommand{\arraystretch}{1.5}
\begin{tabular}{|c|c|c|}
\hline
     Model & Sparticle content & Couplings probed \\ \hline
     $\hbox{Z}_1$ & $ \tilde g, \tilde d_R, \tilde s_R$ & $\lambda^{''}_{uds}, (\delta^d_{RR})_{21}$\\ \hline
     $\hbox{Z}_2$ & $ \tilde g, \tilde d_R, \tilde b_R$ & $\lambda^{''}_{udb}, (\delta^d_{RR})_{31}$\\ \hline
     $\hbox{BM}_1$ & $ \tilde g, \tilde b_R, \tilde b_L, ( \tilde t_L),  \tilde d_L, ( \tilde u_L)$ &
                             $\lambda^{''}_{udb}, (\delta^d_{LL})_{31}, (A_b - \mu \tan\beta)$\\ \hline
     $\hbox{BM}_2$ & $ \tilde g, \tilde b_R,  \tilde d_L, ( \tilde u_L)$ &
                             $\lambda^{''}_{udb}, (\delta^d_{LR})_{31}$\\ \hline
     $\hbox{GS}$ & $ \tilde \chi^\pm, (\tilde \chi^0), \tilde b_R, \tilde b_L, ( \tilde t_L) $ & $\lambda^{''}_{udb}, (A_b - \mu \tan\beta)$\\ \hline
     $\hbox{CK}$ & $ \tilde \chi^\pm, (\tilde \chi^0), \tilde b_R, \tilde t_R, \tilde b_L, ( \tilde t_L) $ & $\lambda^{''}_{tdb}, (A_b - \mu \tan\beta), (A_t - \mu \cot\beta) $\\ \hline
\end{tabular}
\caption{The models considered in this paper. The superpartners in parenthesis do not contribute to the oscillation process but are required by  $SU(2)_L$ gauge invariance. All other sparticles are decoupled and all other RPV or FV couplings are set to zero. All squarks are assumed to be mass degenerate.}
\label{tab:modellist}
\end{center}
\end{table}
We finally come to the discussion of the various contributions to $n-\bar{n}$ oscillations that can arise in BRPV supersymmetry and compare their sensitivity to the previous constraints.
Our philosophy is as follows.

In the spirit of simplified models, we always test one RPV coupling at the time setting all the remaining to zero.
For each process we consider a simplified spectrum where all the particles not contributing to  the actual $n-\bar n$ diagram are assumed to be decoupled, i.e.~taken to be very heavy. The constraints from the other physical processes discussed in Section~4 will be applied to such model.
The only important exception to the above rule arises when some superpartners belong to a multiplet of $SU(2)_L$.
It is then necessary, because of $SU(2)_L$ gauge invariance, to assume the other member of the doublet to be present in the spectrum as well, and nearly degenerate in mass. This case arises when LH squarks or a Wino-like chargino are present in the diagrams.
As far as the spectrum is concerned, we will always consider all the relevant squarks as degenerate and scale their production cross-section accordingly.  

We separate between strong and electroweak contributions. In the strong processes, the only superpartners present in the spectrum are the relevant squarks and the gluino $\tilde g$. Similarly, the electroweak contribution will be computed for models with only squarks and one Wino-like chargino $\tilde \chi^\pm$ (and the corresponding neutralino). There is a large number of possible processes available but, when comparing contributions amongst themselves and particularly against the bounds from di-nucleon decay, we reduce the list to what is shown in Table~\ref{tab:modellist}.

 \subsection{Strong contributions}

\begin{figure}[t]
\begin{center}
\includegraphics[width=1.\textwidth]{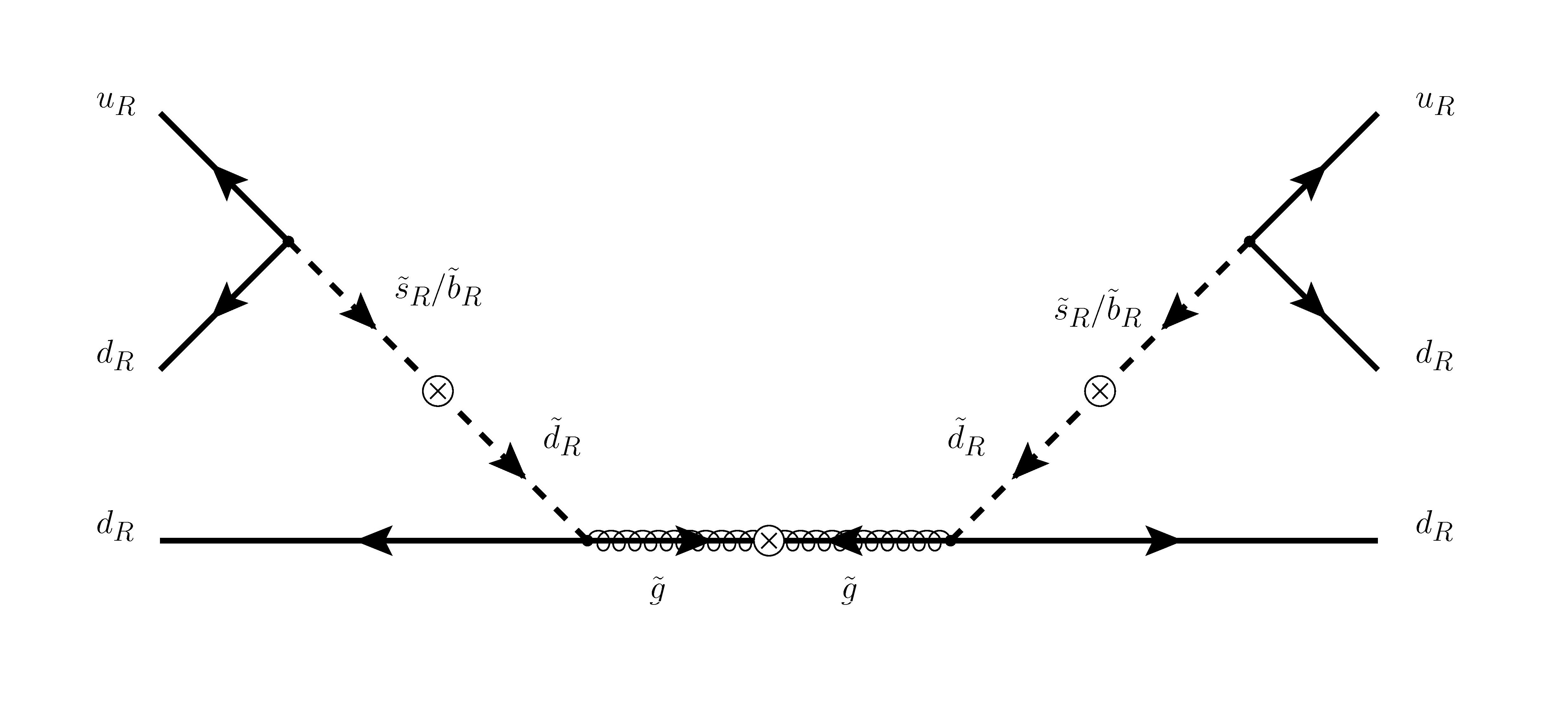}
\caption{The Zwirner diagram contributing to Models Z${}_1$ and Z${}_2$. \label{fig:Zwirner}}
\end{center}
\end{figure}

The first SUSY contribution to $n-\bar{n}$ oscillations that we consider was  first discussed by Zwirner in ref.~\cite{Zwirner:1984is}, involving the Feynman diagram shown in Fig.~\ref{fig:Zwirner}, and gives rise to the operator,
\begin{equation}
\label{Zwirnerop}
\mathcal{L}^{Z}_{n\bar{n}}=C_{n\bar{n}}^{Z}~ \urdrdr + \mathrm{h.c.},
\end{equation}

Here we have a choice between using a RH strange squark or a RH bottom squarks in the diagram, probing separately the two RPV couplings  $\lambda_{uds}^{''}$ and  $\lambda_{udb}^{''}$. We will consider both cases, although the first one is seriously constrained by di-nucleon decays to Kaons.

The coefficient has the following form:
\begin{eqnarray}
C_{n\bar{n}}^{Z} &=& \frac{4}{3}\frac{g_s^2}{m_{\tilde{g}}}\left|
\frac{\lambda_{udk}^{''} (\delta^d_{RR})_{k1}}{\overline{m}^2_{D}}\right|^2,
\label{eq:CZ}
\end{eqnarray}
where $k=s\mbox{ or }b$ and we employed the mass-insertion approximation as defined in Eq.~(\ref{eq:deltas}), assuming nearly-degenerate RH  squarks:
$m_{\tilde{d}_R}=m_{\tilde{s}_R}=\overline{m}_{D}$ in one case and $m_{\tilde{d}_R}=m_{\tilde{b}_R}=\overline{m}_{D}$ in the other. 
The $n-\bar{n}$ oscillation time, arising from the contribution \eqref{Zwirnerop}, is then,
\begin{equation}
\tau_{n\bar{n}}  = \frac{1}{C_{n\bar{n}}^{Z} \langle \bar{n} | \urdrdr | n \rangle },
\end{equation}
 Numerically we obtain:
\begin{equation}
\tau_{n\bar{n}}  = (2.5 \times 10^8 ~\mathrm{s})\times
\frac{m_{\tilde g}}{1.2~\mathrm{TeV}}
\left(\frac{\overline{m}_{D}}{500~\mathrm{GeV}}\right)^4
\left( \frac{2\times10^{-6}}{\lambda^{''}_{udk}}\right)^2
\left( \frac{0.01}{(\delta^d_{RR})_{k1}}\right)^2
 \frac{(250~\mathrm{MeV})^6}{\langle \bar{n} | \urdrdr | n \rangle}
 \label{eq:CZnum}
\end{equation}
The above value of the oscillation time is at the level of the present indirect bound, $\tau_{n\bar{n}}^{\mathrm{exp}} > 2.7{\times}10^{8}$\,s~\cite{Abe:2011ky}.
The values of $\langle \bar{n} | \urdrdr | n \rangle$ reported in the literature vary by more
than one order of magnitude: here we adopted the estimate employed in \cite{Csaki:2012zr}.
Note that a bound set by $n-\bar{n}$ on $\lambda^{''}$ will vary as the square root of $\langle \bar{n} | \urdrdr | n \rangle$.

Finally, we stress that the above contributions require flavour violation beyond
minimal flavour violation (MFV) \cite{D'Ambrosio:2002ex}. In fact, under the MFV hypothesis, the right-handed squarks are diagonal in flavour space, one would have $(\delta^d_{RR})_{k1}=0$, hence the Zwirner contribution would vanish.

\begin{figure}[t]
\begin{center}
\includegraphics[width=1.\textwidth]{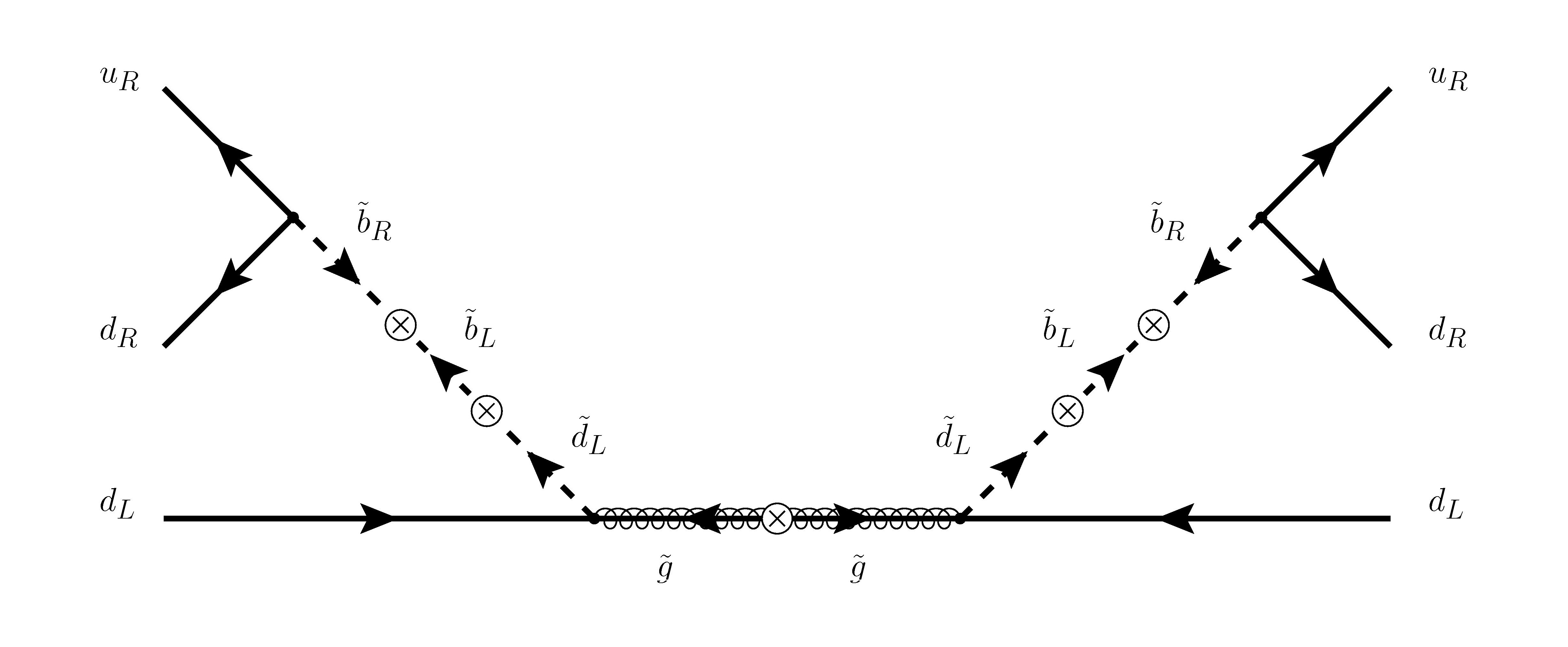}
\caption{The Barbieri and Masiero diagram contributing to BM${}_1$. \label{fig:BM}}
\end{center}
\end{figure}

One way to get a non-vanishing tree level contribution to $n-\bar{n}$  under the assumption of MFV is to mix the $\tilde{s}_R/\tilde{b}_R$ with their left-handed counterparts $\tilde{s}_L/\tilde{b}_L$, which can be done by inserting the corresponding off-diagonal mass mixing element $m_{s/b}\left( A_{s/b} -\mu \tan\beta \right)$. Since MFV allows $\tilde{s}_L/\tilde{b}_L$ to mix with the first generation left-handed squark $\tilde{d}_L$, we can have the diagram in Fig.~\ref{fig:BM}, where two of the external quarks are now taken to be left-handed. A similar contribution was pointed out by Barbieri and Masiero in ref.~\cite{Barbieri:1985ty}. Here we use the explicit expression of the LR-mixing in terms of the RPV-MSSM parameters.
As discussed in Appendix A, the fact that two of the external down-type quarks are left-handed implies that, above the EWSB scale, the corresponding operator is of dimension 11, since it involves two Higgs fields that contracts these two left-handed external quarks. Below the EWSB scale, the two Higgs VEVs combine with the corresponding Yukawa couplings that enter in the SUSY breaking couplings between the left- and right-handed squarks and the external Higgs fields, and make up the factor of $m_{s/b}$ that appears in the off-diagonal mass mixing insertion.

The fact that these contributions are proportional to  $m_{s/b}^2$ implies that the contribution from the s-strange is less important than the contribution from the sbottom. As a consequence, since one needs two left-right mixing insertions, as well as  two flavour insertions,
the contribution from the s-strange is negligible compared to the constraint coming from di-nucleon decay. Note that, as will be discussed below,  di-nucleon decay constrains  $\lambda_{uds}^{''}$ much more than $\lambda_{udb}^{''}$.
Therefore, we focus only on the sbottom contribution, which involves only $\lambda_{udb}^{''}$.

\begin{figure}[t]
\begin{center}
\includegraphics[width=1.\textwidth]{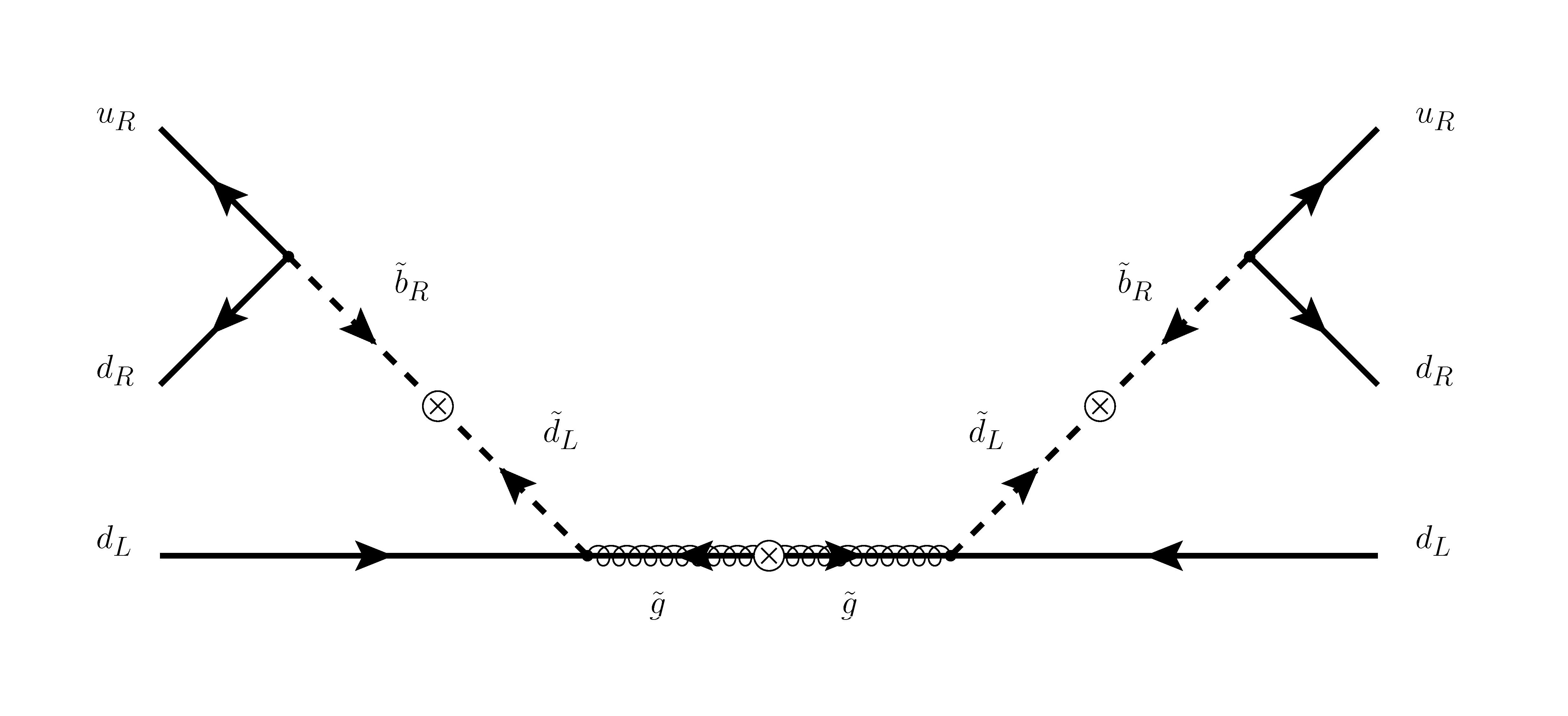}
\caption{The Barbieri and Masiero diagram, in presence of flavour-violating A-terms, contributing to Model BM${}_2$.\label{fig:BM1in}}
\end{center}
\end{figure}

Below the EWSB scale, the dimension 11 operator becomes the following dimension 9 operator,
\begin{equation}
\mathcal{L}_{n\bar{n}}=C_{n\bar{n}}^{\mathrm{BM}} \, \urdrdl+ \mathrm{h.c.},
\end{equation}
where $\urdrdl$ can be found in \eqref{relevantops} and
\begin{equation}
\label{eq:BM}
C_{n\bar{n}}^{\mathrm{BM}_1}  =  \frac{4}{3} \,\frac{g_s^2\,(\lambda_{udb}^{''})^2\,m_b^2 \,(A_b - \mu\tan\beta)^2 \,\left|(\delta^d_{LL})_{31}\right|^2}{m_{\tilde{b}_R}^4\,m_{\tilde{b}_L}^4\,m_{\tilde{g}}}.
\end{equation}
Similarly, in presence of off-diagonal entries in the A-term matrix, flavour violation and the chirality flip can both be obtained by a single mass insertion as shown in Fig.~\ref{fig:BM1in}, yielding
\begin{equation}
\label{eq:BM-LR}
C_{n\bar{n}}^{\mathrm{BM}_2}  =  \frac{4}{3} \,\frac{g_s^2\,(\lambda_{udb}^{''})^2\, \left|(\delta^d_{LR})_{13}\right|^2}{m_{\tilde{b}_R}^2\,\,m_{\tilde{d}_L}^2\,m_{\tilde{g}}}.
\end{equation}
For instance, the numerical result for the ${\mathrm{BM}}_1$ contribution is:
\begin{align}
\tau_{n\bar{n}}^{\mathrm{BM}_1}  = ~& (2.5 \times 10^8 ~\mathrm{s})\times
\frac{m_{\tilde g}}{1.8~\mathrm{TeV}}
\left(\frac{m_{\tilde{b}_R}}{1.1~\mathrm{TeV}}\right)^4
\left(\frac{m_{\tilde{b}_L}}{1.1~\mathrm{TeV}}\right)^4
\left( \frac{50~\mathrm{TeV}}{A_b - \mu\tan\beta}\right)^2 \times \nonumber\\
&\left( \frac{2\times 10^{-5}}{\lambda^{''}_{udb}}\right)^2
\left( \frac{0.05}{(\delta^d_{LL})_{31}}\right)^2
 \frac{(250~\mathrm{MeV})^6}{\langle \bar{n} | \urdrdl | n \rangle}.
\end{align}

As was mentioned at the beginning of the section, we now present our results within a set of simplified models that feature only the particle content relevant for the above $n-\bar{n}$ diagrams.
 We further classify according to the source of flavour violation when relevant.  The models are summarised in Tab.~\ref{tab:modellist}.
 \subsection*{Model  $\hbox{Z}_1$, spectrum $ \tilde g, \tilde d_R, \tilde s_R$, couplings $\lambda^{''}_{uds}, (\delta^d_{RR})_{21}$}
 \begin{figure}[t]
\begin{center}
\includegraphics[width=0.4\textwidth]{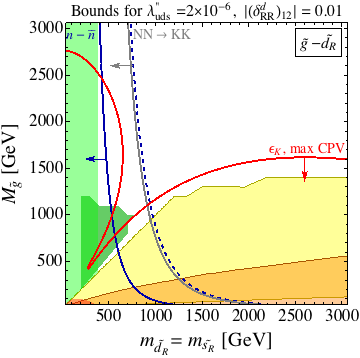}
\hspace{0.4cm}
\includegraphics[width=0.4\textwidth]{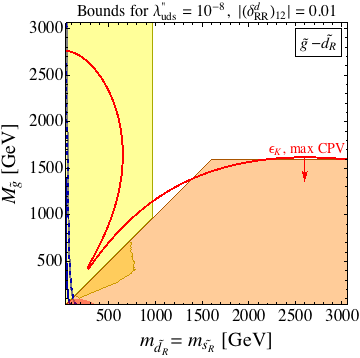}
\includegraphics[width=0.4\textwidth]{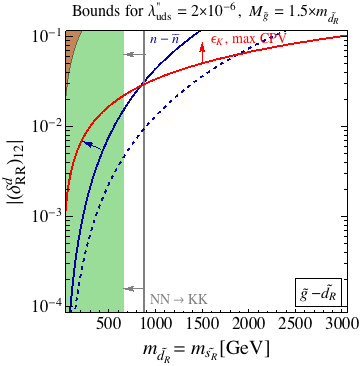}
\hspace{0.4cm}
\includegraphics[width=0.4\textwidth]{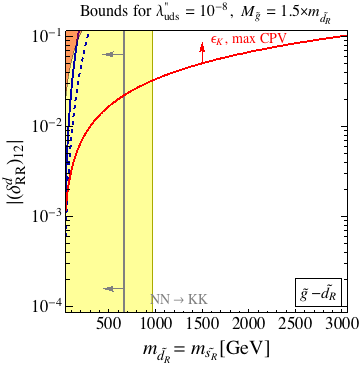}
\includegraphics[width=0.4\textwidth]{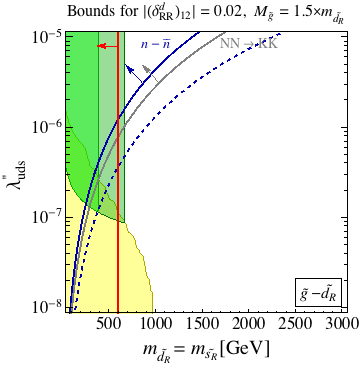}
\hspace{0.4cm}
\includegraphics[width=0.4\textwidth]{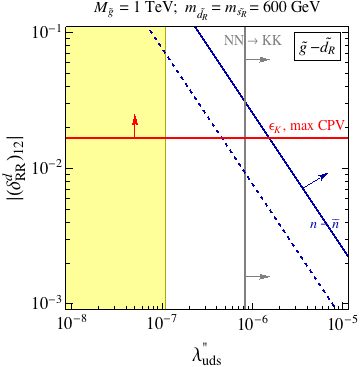}
\caption{Bounds and $n-\bar{n}$ prospects for the Model $\hbox{Z}_1$ for different choices of the parameters.
The low energy constraints are represented as follows.
Red regions: $\Delta m_K$. Red lines:  $\epsilon_K$. Gray lines: $NN{\to}KK$. Blue lines: $n-\bar{n}$. Dashed blue lines: prospected sensitivity of the $n-\bar{n}$ ESS experiment. The LHC constraints are shown as follows. Light green regions: CMS dijet \cite{Khachatryan:2014lpa}. Dark green regions: ATLAS multijet \cite{Aad:2015lea}. Yellow regions: displaced jets \cite{Liu:2015bma,Csaki:2015uza}. Orange regions: CMS long-lived particles \cite{CMS:2015kdx}. See the text for details. \label{fig:lambda_uds}}
\end{center}
\end{figure}
In the presence of only gluinos and the RH down-type squarks $\tilde{d}_R$ and $\tilde{s}_R$ (that in the following we are going to assume almost degenerate),
$n-\bar{n}$ oscillations can occur via the diagram of Fig.~\ref{fig:Zwirner}.
In order for the diagram not to vanish, flavour violation is required either in the 1-2 or in the 1-3 sector. In other words, RH down squarks have to mix either with strange or bottom squarks. Here we consider the first case, while the second one will be presented in the next subsection.
As previously discussed in section \ref{sec:FCNC}, flavour violation in the 1-2 sector gives rise to contributions to $K-\bar{K}$ mixing that are stringently constrained by the observed Kaon mass splitting $\Delta m_K$ and CP violation parameter $\epsilon_K$, see Fig.~\ref{fig:delta12bound}.
As explained in section \ref{sec:NN}, the RPV coupling $\lambda_{uds}^{''}$ that controls $n-\bar{n}$  oscillation within this model is also constrained
by non-observation of di-nucleon decays.

We can now display the above constraints together with the bound from $n-\bar{n}$ oscillation and the ESS facility potential.
These are shown in Fig.~\ref{fig:lambda_uds} for different choices of the parameters.
In the figure, we display the bound imposed by $\Delta m_K$ as red regions, while red lines correspond to the constraint that $\epsilon_K$ would give
in presence of a maximal ($\pi/4$) CPV phase of $(\delta^d_{RR})_{12}$.
 The blue lines depict the present bound from $n-\bar{n}$ oscillations ($\tau_{n\bar{n}}^{\mathrm{exp}} > 2.7{\times}10^{8}$\,s),
 setting $\langle \bar{n} | \urdrdr| n \rangle = (250\mbox{ MeV})^6$.
The dashed blue lines are the bounds that will be reached if a new experiment would have
a sensitivity up to  $\tau_{n\bar{n}}^{\mathrm{exp}} = 3{\times}10^{9}$\,s.
Indeed the proposed experiment at ESS is supposed to improve the sensitivity to the oscillation probability with respect to the ILL-Grenoble experiment by a factor of 1000, which means a factor of 32 in the  oscillation time \cite{EOInnbar}.
 The di-nucleon decay constraint, $\tau_{NN{\to}KK}^{\mathrm{exp}} > 1.7{\times}10^{32}$ years, is shown as gray lines, taking
$\langle NN|\urdrsr|KK\rangle = (150\mbox{ MeV})^5$.

The limits set by LHC searches for new physics are shown in Fig.~\ref{fig:lambda_uds} as follows: the light green regions correspond to the dijet pair search by CMS \cite{Khachatryan:2014lpa}, the dark green regions to our recast of the ATLAS multijet search \cite{Aad:2015lea},
the yellow regions to the limit from displaced jet searches as obtained by \cite{Liu:2015bma,Csaki:2015uza}, the orange regions
are the limits from the recent $\sqrt{s}=13$ TeV CMS search for long-lived particles \cite{CMS:2015kdx}. For further details about the present status of the relevant LHC searches, cf.~section \ref{sec:LHC}.

Consistently with the life-times displaced in Fig.~\ref{fig:decay-lenghts}, we see that for $\lambda_{uds}^{''} \gtrsim 10^{-7}$ squarks have prompt decays even if lighter than gluinos (cf.~the left panel of the third row), such that multijet (dark green) and dijet pairs (light green) searches set the most relevant LHC bounds on the half-plane $m_{\tilde{d}_R} < M_{\tilde g}$: this is shown in the upper-left panel of the figure, corresponding to $\lambda_{uds}^{''} =2\times10^{-6}$. On the other hand, gluinos lighter than squarks mostly decay to displaced jets. This is why the limit of  \cite{Liu:2015bma,Csaki:2015uza} (yellow region) dominates for $m_{\tilde{d}_R} > M_{\tilde g}$. Decreasing the RPV coupling below that level makes all particles decaying more slowly: this is shown in the upper-right panel of the figure where $\lambda_{uds}^{''} =10^{-8}$. The dominant bounds come from searches for displaced jets for $m_{\tilde{d}_R} < M_{\tilde g}$ and long-lived R-hadrons
for $m_{\tilde{d}_R} > M_{\tilde g}$. This latter bound is given by the recent 13 TeV search performed by CMS \cite{CMS:2015kdx} and -- in terms of reach in SUSY masses -- is the strongest to date among those relevant for us, corresponding to $M_{\tilde g}\gtrsim 1.6$ TeV.

\begin{figure}[t]
\begin{center}
\includegraphics[width=0.45\textwidth]{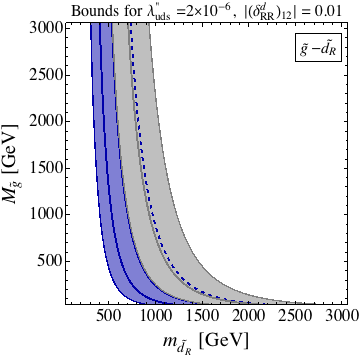}
\caption{Impact of hadronic uncertainties on the bounds $n-\bar{n}$ (blue region) and $NN\to KK$ (gray regions). See the text for details.
\label{fig:hadronic-unc}}
\end{center}
\end{figure}

In Fig.~\ref{fig:lambda_uds}, we have fixed  $\langle \bar{n} | \urdrdr | n \rangle$ and $\langle NN|\urdrsr|KK\rangle$ to the above values for illustration purposes, as in \cite{Csaki:2012zr}. In Fig.~\ref{fig:hadronic-unc}, we depict the uncertainty due to the hadronic matrix elements: the blue band correspond
to the present $n-\bar{n}$ bound taking $(1/3)\times  (250\mbox{ MeV})^6  \le \langle n|\urdrdr|\bar{n}\rangle \le 3\times (250\mbox{ MeV})^6 $. The gray band corresponds to one order of magnitude variation of the matrix elements of the di-nucleon decay as in the left panel of Fig.~\ref{fig:NNKK2}.
As in Fig.~\ref{fig:lambda_uds}, the blue dashed line corresponds to the sensitivity of the ESS experiment with
$ \langle n|\urdrdr|\bar{n}\rangle = (250\mbox{ MeV})^6 $. 
Note that the hadronic uncertainties affect more the bounds on superpartner masses in the case of $NN\to KK$, as the di-nucleon decay rate scales quadratically with the matrix element, while the neutron oscillation time scales linearly.

From Figs.~\ref{fig:lambda_uds} and $\ref{fig:hadronic-unc}$, we see that the stringent bounds set by the di-nucleon decay
tend to be stronger than $n-\bar{n}$ in constraining the parameter space.
Remarkably, the planned improvement in the sensitivity to $n-\bar{n}$ oscillations might however
-- depending on the hadronic matrix elements, as well as on the value of $\left( \delta^d_{RR}\right)_{12}$ -- explore new territories even in this unfavorable case. 
 \subsection*{Model   $\hbox{Z}_2$, spectrum $ \tilde g, \tilde d_R, \tilde b_R$, couplings $\lambda^{''}_{udb}, (\delta^d_{RR})_{31}$}
\begin{figure}[t]
\begin{center}
\includegraphics[width=0.4\textwidth]{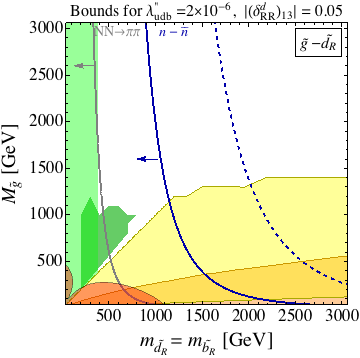}
\hspace{0.4cm}
\includegraphics[width=0.4\textwidth]{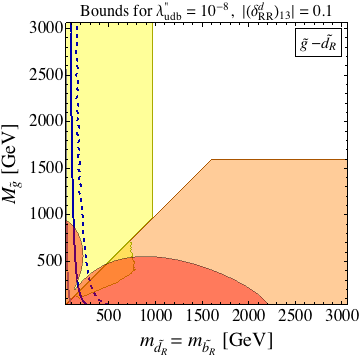}
\includegraphics[width=0.4\textwidth]{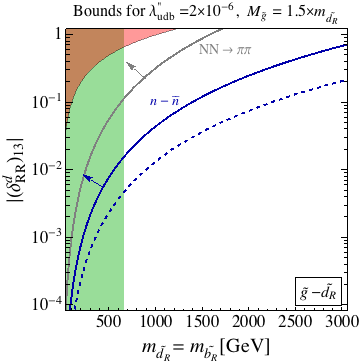}
\hspace{0.4cm}
\includegraphics[width=0.4\textwidth]{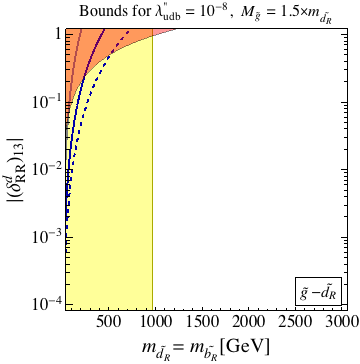}
\includegraphics[width=0.4\textwidth]{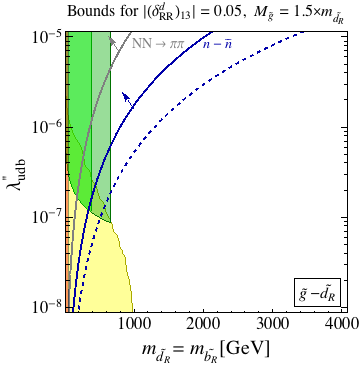}
\hspace{0.4cm}
\includegraphics[width=0.4\textwidth]{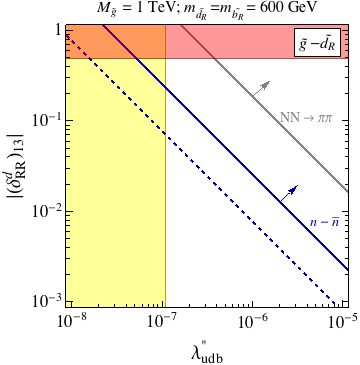}
\caption{Bounds and $n-\bar{n}$ prospects for
the Model $\hbox{Z}_2$ for different choices of the parameters. The low energy constraints are represented as follows.
Red regions: $\Delta m_B$. Gray lines: $NN{\to}\pi\pi$. Blue lines: $n-\bar{n}$. Dashed blue lines: prospected sensitivity of the $n-\bar{n}$ ESS experiment. The LHC constraints are shown as follows. Light green regions: CMS dijet \cite{Khachatryan:2014lpa}. Dark green regions: ATLAS multijet \cite{Aad:2015lea}. Yellow regions: displaced jets \cite{Liu:2015bma,Csaki:2015uza}. Orange regions: CMS long-lived particles \cite{CMS:2015kdx}. See the text for details.\label{fig:ludb-d13}}
\end{center}
\end{figure}
Model $\hbox{Z}_2$ concerns the contribution in Fig.~\ref{fig:Zwirner}  
for the case where the internal squark is a sbottom instead of a s-strange. 
In this case, flavour violation occurs in the 1-3 sector where the constraints (coming from $B-\bar{B}$ mixing)
on $\left( \delta^d_{RR}\right)_{13}$
are much milder than the analogous ones in the 1-2 sector, as shown in the upper-right plot of Fig.~\ref{fig:delta12bound}.
The simplified model we are going to study for this case only involves RH down and bottom squarks ($\tilde{d}_R$ and $\tilde{b}_R$) and
gluinos.

Furthermore, unlike the previous case, there are no relevant bounds on $\lambda_{udb}^{''}$ from $NN\to KK$
stronger than $n-\bar{n}$ itself. As discussed in section \ref{sec:NN}, the other di-nucleon decay mode $NN\to \pi\pi$ is possibly relevant.
However, it turns out to give a subdominant constraint, barring conspiracies of the hadronic matrix elements.
This makes this scenario particularly suitable to accommodate $n-\bar{n}$ oscillations at the level of the present experimental sensitivity.
We summarise the experimental situation in Fig.~\ref{fig:ludb-d13}, where the colour code is as in the previous subsection.
The only difference is given by the red regions, which now depict bounds from $\Delta m_B$ (the $B-\bar{B}$ mixing CPV observables have an equivalent impact even with maximal CPV phases), and the gray lines which correspond to the limit $\tau_{nn\to \pi^0\pi^0} >4.04{\times}10^{32}$ years, calculated choosing $\langle NN|\urdrdr|\pi\pi\rangle =(250\mbox{ MeV})^5$.
As we can see, the experiment proposed at ESS can give a spectacular improvement in the sensitivity.
In particular, we see that multi-TeV squarks might still induce observable oscillation rates (cf.~the left panels
in the second and third rows of Fig.~\ref{fig:ludb-d13}), arguably beyond the reach of the LHC.
On the other hand, small amounts of RPV, $\lambda_{udb}^{''} \lesssim 10^{-7}$,
make any low-energy process irrelevant, leaving direct collider searches as the privileged way to test this kind of models. This is depicted by the plots in the third row of Fig.~\ref{fig:ludb-d13}.
 \subsection*{Model      $\hbox{BM}_1$, spectrum $ \tilde g, \tilde b_R, \tilde b_L, ( \tilde t_L),  \tilde d_L, ( \tilde u_L)$, couplings
                             $\lambda^{''}_{udb}, (\delta^d_{LL})_{31}, (A_b - \mu \tan\beta)$}
We turn now to consider a model with no flavour mixing among RH squarks (as predicted by MFV scenarios).
The flavour transition necessary to generate a $\Delta B=2$ operator via the $\lambda^{''}$ couplings
can then occur in the LH squark sector and be transmitted to the RH sector through LR squark mixing, see Fig.~\ref{fig:BM}.
The minimal particle content required to give rise to this contribution consists of gluinos and down squarks both of RH and LH kinds.
As a consequence of the squark chirality flip, the resulting oscillation probability depends on the relevant down quark mass. Diagrams involving
sbottoms are then enhanced by a factor $(m_b/m_s)^2$ compared to those featuring strange squarks, hence they are the only ones
of possible phenomenological relevance. Neutron oscillation are then controlled by $\lambda^{''}_{udb}$ and  $(\delta^d_{LL})_{13}$.
The particle content is given by $\tilde{b}_R$, $\tilde{b}_L$ and $\tilde{d}_L$ (and thus $\tilde{t}_L$ and $\tilde{u}_L$ too).

The most stringent flavour constraints on this scenario come from $b\to d\gamma$ transitions,
due to sizeable contributions to flavour violating dipole operators induced by the large LR mixing. The corresponding bound for the illustrative case
of $\mu\times\tan\beta=10$ TeV is shown in the lower-left panel of Fig.~\ref{fig:delta12bound}.
\begin{figure}[t]
\begin{center}
\includegraphics[width=0.4\textwidth]{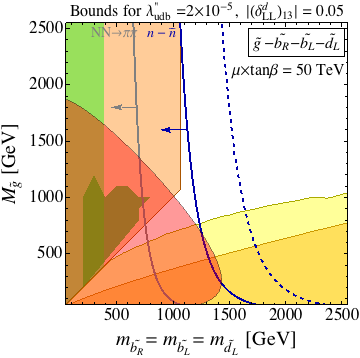}
\hspace{0.4cm}
\includegraphics[width=0.4\textwidth]{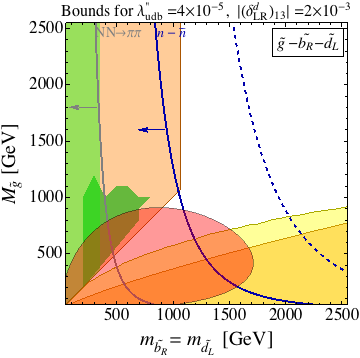}
\caption{Bounds and $n-\bar{n}$ prospects for the Models $\hbox{BM}_1$ (left) and $\hbox{BM}_2$ (right). The red region corresponds to the $b\to d\gamma$ bound. The other constraints are represented as in Fig.~\ref{fig:ludb-d13}. \label{fig:BM-ludb-d13}}
\end{center}
\end{figure}
In the right panel of Fig.~\ref{fig:BM-ludb-d13}, we show the $b\to d\gamma$ constraint (as a red region) together with the other constraints (colour code as in the previous subsections), for an illustrative choice of the parameters. Notice that given the presence of long-lived $\tilde{u}_L$, $\tilde{d}_L$ and  $\tilde{t}_L$ the dominant LHC constraint come from searches for long-lived particles, also in the part of the plane where squarks are lighter than gluinos, and relatively large RPV couplings, $\lambda^{''}_{udb} = \mathcal{O}(10^{-5})$. Still, searches for $n-\bar{n}$ oscillation have the potential of going beyond the LHC in testing the parameter space of this model.
\subsection*{Model $\hbox{BM}_2$, spectrum $ \tilde g, \tilde b_R,  \tilde d_L, ( \tilde u_L)$, couplings  $\lambda^{''}_{udb}, (\delta^d_{LR})_{31}$}
In the model discussed above, where both LH and RH squarks are present, flavour violation can also occur through a flavour off-diagonal A-term.
The diagram leading to  $n-\bar{n}$ oscillation is as in Fig.~\ref{fig:BM1in}, with the flavour and the LR mixing being simultaneously provided by a single
mass insertion. The resulting contribution is given by Eq.~(\ref{eq:BM-LR}): the corresponding constraints are shown in the right plot of
Fig.~\ref{fig:BM-ludb-d13}.
Flavour mixing in the LR sector gives a large contribution to the dipole transition responsible of  $b\to d\gamma$ and is therefore tightly constrained,
as we can see in the lower-right panel of Fig.~\ref{fig:delta12bound}.
Relatively larger values of $\lambda^{''}_{udb}$ than in the $(\delta^d_{LL})_{13}$ case are then needed to have a signal of $n-\bar{n}$ oscillation
without too large flavour violation. This can be seen by comparing the two plots of Fig.~\ref{fig:BM-ludb-d13}.

\subsection{Electroweak contributions}
\begin{figure}[t]
\begin{center}
\includegraphics[width=0.9 \textwidth]{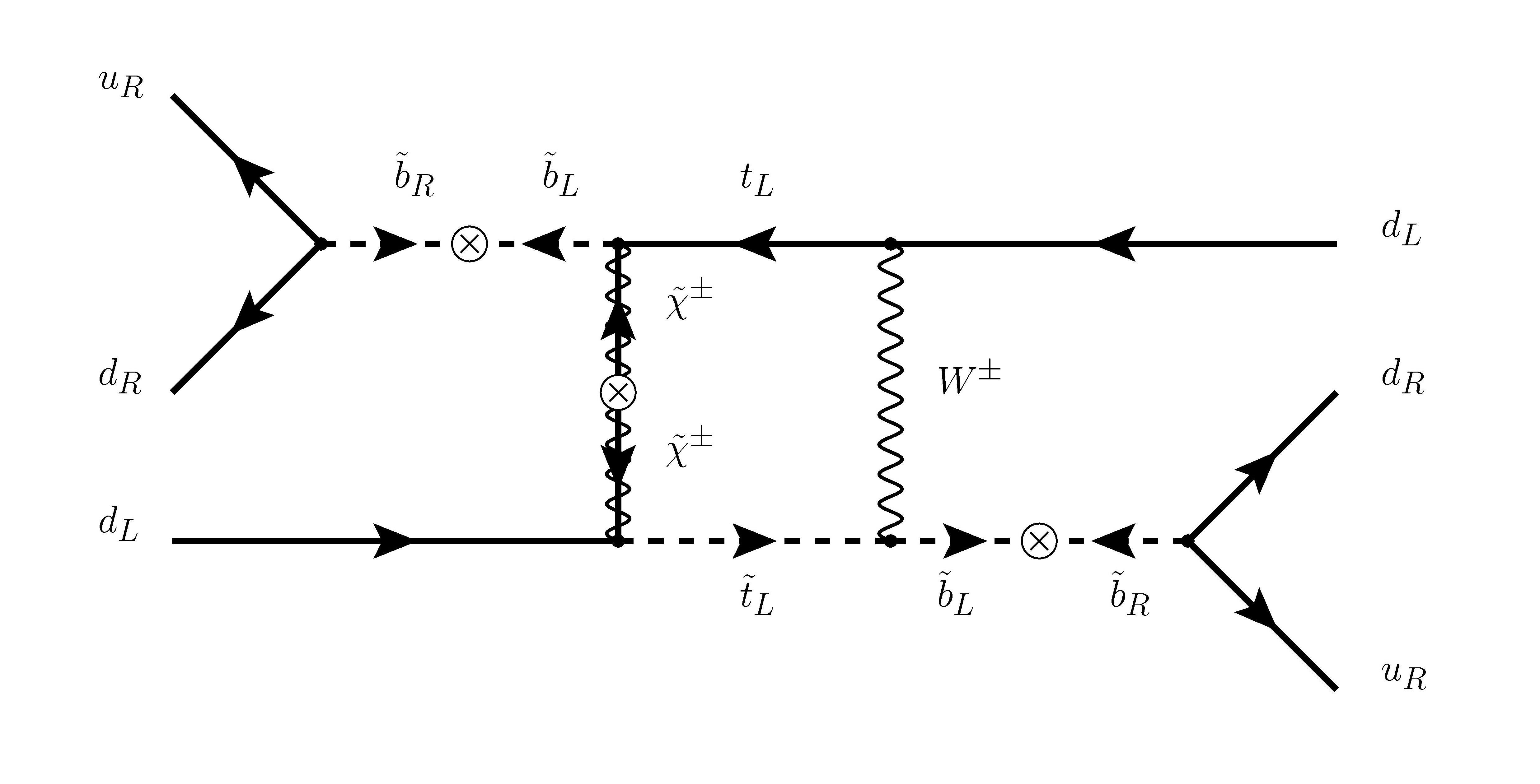}
\caption{The Goity and Sher diagram contribution to Model GS. \label{fig:GS}}
\end{center}
\end{figure}
\def\bstname{fdp}

All the above oscillation mechanisms rely on the presence of a gluino in the diagram.
If the gluino is decoupled from the theory, it is still possible to use charginos to construct {\it electroweak} SUSY contributions to $n-\bar{n}$ oscillations.
Since the chargino does not carry colour degrees of freedom, these will necessarily be loop contributions. One possibility, originally proposed by Goity and Sher~\cite{Goity:1994dq}, involves a flavour changing box diagram, shown in Fig.~\ref{fig:GS}, which is essentially the supersymmetrization of the famous GIM diagram \cite{Glashow:1970gm}. The presence of a Wino-like chargino and a $W$ also means that we must necessarily include some LH squarks in the model.

Even in this case we have various options for the choice of which squarks to retain in our simplified model. The choice between $\tilde s_R$ and $\tilde b_R$ is clear and already explained in the previous sections: we choose $\tilde b_R$ since the $\tilde b_L-\tilde b_R$ mixing is proportional to the mass of the $b$-quark instead of that of the $s$-quarks, as well as because the coupling $\lambda^{''}_{udb}$ is much less constrained by di-nucleon decay.
Once we have chosen to introduce a $\tilde b_L$ in the spectrum, $SU(2)_L$ gauge invariance requires us to include the LH stop $\tilde t_L$ as well.
Minimality thus suggests to use the LH stop in the FV box diagram and decouple the $\tilde u_L$ and $\tilde c_L$ quarks. Indeed, some splitting between the masses of the LH $u$-type squarks is required in order for the box diagram not to vanish due to the unitarity of the CKM matrix. The final diagram and the non decoupled field content is shown in Fig.~\ref{fig:GS}.

\begin{figure}[t]
\begin{center}
\includegraphics[width=0.6\textwidth]{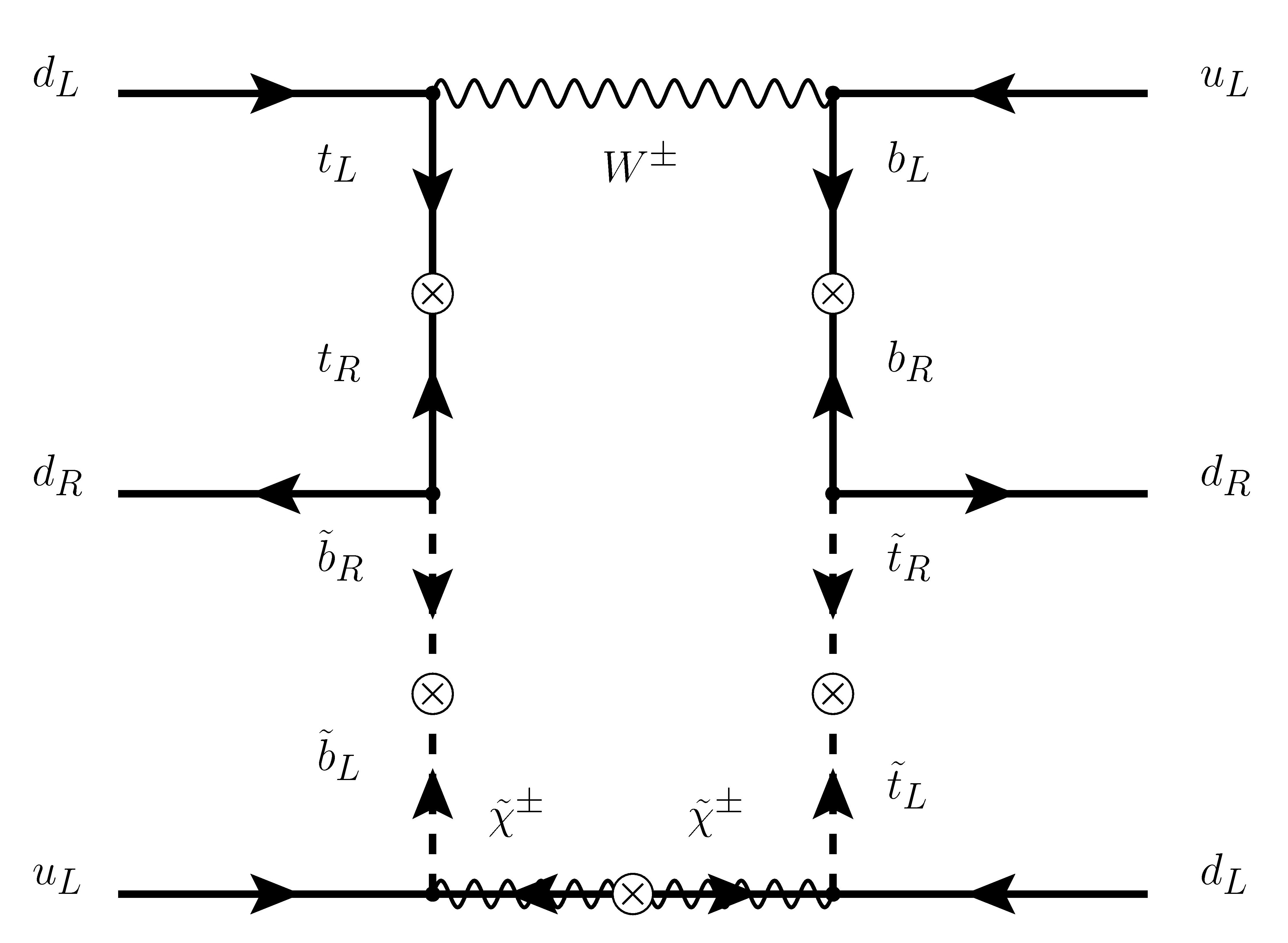}
\caption{The Chang and Keung diagram contributing to Model CK. \label{fig:CandK}}
\end{center}
\end{figure}

An alternative possibility, proposed be Chang and Keung~\cite{Chang:1996sw} and shown in Fig.~\ref{fig:CandK}, is to have the RPV vertex appear inside the loop. This is the only case where we can have a $u_L$ quark appearing in the effective operator, which is in fact $\uldldr$, the Parity conjugate of the previous $\urdrdl$. As for the choice of the internal quarks/squarks, the largest contribution comes from the third family, as shown in Fig.~\ref{fig:CandK}. This is thus the only case that is sensitive to $\lambda^{''}_{tds}$, which is a coupling of great interest in collider searches.

In the case of Fig.~\ref{fig:GS}, one obtains~\cite{Goity:1994dq},
\begin{equation}
\label{opGS}
\mathcal{L}_{n\bar{n}}=C_{n\bar{n}}^{\mathrm{GS}} \,\urdrdl + \mathrm{h.c.}
\end{equation}
where the 
\begin{equation}
\label{GSgeneral}
C_{n\bar{n}}^{\mathrm{GS}}  = \frac{g^4(\lambda_{udb}^{''})^2\,m_b^2 \,(A_b - \mu\tan\beta)^2 \, m_{\tilde{\chi}^{\pm}}   }{32 \,\pi^2\,m_{\tilde{b}_R}^4\,m_{\tilde{b}_L}^4}
\xi_{jj'} J( m_{\tilde{\chi}^\pm}^{2} ,m_W^2, m_{u_{j} }^2, m_{\tilde{u}_{j'} }^2 ),
\end{equation}
and
\begin{equation}
\label{Jlog}
J(x_1,x_2,x_3,x_4)=\sum_{i=1}^4 \frac{x_i^2 \log (x_i)}{\prod_{k\neq i} (x_i - x_k)}.
\end{equation}
$\xi_{jj'}$ is a combination of CKM matrix elements:  $\xi_{jj'} = V^\dagger_{1j} V_{j3} V^\dagger_{1j'} V_{j'3}$. Since the simplified model we consider involves only the third family up-type squark, we set $j'=3$ in the general formula \eqref{GSgeneral}, implying that the third family up-type quark ($j=3$) gives the leading contribution, as is indicated by Fig.~\ref{fig:GS}. Note that the function $J$ in \eqref{Jlog} does not depend on a reference scale as the sum vanishes for any constant value inside the logarithm.  

The contribution of the diagram in Fig.~\ref{fig:CandK} is instead given by~\cite{Chang:1996sw}
\begin{equation}
\label{opCK}
\mathcal{L}_{n\bar{n}}=C_{n\bar{n}}^{\mathrm{CK}} \,\uldldr  + \mathrm{h.c.}
\end{equation}
where
\begin{align}
\label{coeffCK}
C_{n\bar{n}}^{\mathrm{CK}}  =~& \frac{g^4}{16\pi^2}(\lambda^{''}_{tdb})^2 m_{\tilde \chi^\pm} m_t^2 m_b^2 (A_b - \mu \tan\beta)(A_t - \mu \cot\beta)
(V_{td}V^*_{ub})^2 \times\nonumber\\
& I(m_{\tilde \chi^\pm}^2, m_W^2, m_t^2, m_b^2, m_{\tilde t}^2, m_{\tilde b}^2 ),
\end{align}
\begin{eqnarray}
\label{Ifunction}
I(x_1,x_2,x_3,x_4,x_5,x_6)&=&\int_0^\infty\frac{x dx}{(x+x_5)^2(x+x_6)^2\prod_{k=1}^4(x+x_k)}\\
&=&\frac{\partial^2}{\partial x_5\partial x_6}\sum_{i=1}^6\frac{x_i\log x_i}{\prod_{k\neq i}(x_k - x_i)}.\nn
\end{eqnarray} 
We now discuss the exclusion regions for these two electroweak models in turn.

\subsection*{Model $\hbox{GS}$, spectrum $ \tilde \chi^\pm, \tilde b_R, \tilde b_L, ( \tilde t_L) $, couplings $\lambda^{''}_{udb}, (A_b - \mu \tan\beta)$}
\begin{figure}[t]
\begin{center}
\includegraphics[width=0.4\textwidth]{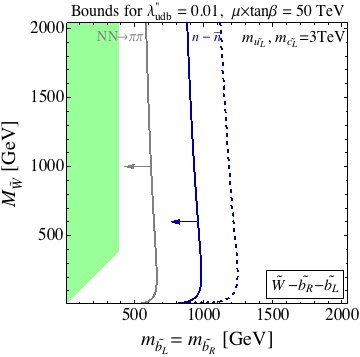}
\hspace{0.4cm}
\includegraphics[width=0.4\textwidth]{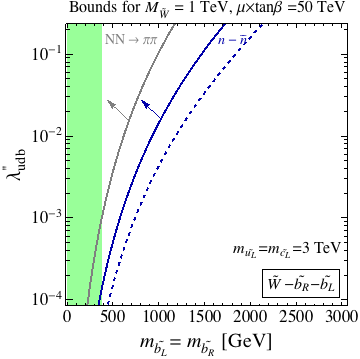}
\caption{Bounds and $n-\bar{n}$ prospects for
the Model $\hbox{GS}$. The colour code is as in Fig.~\ref{fig:ludb-d13}. \label{fig:GSplots}}
\end{center}
\end{figure}
The results for the model of Fig.~\ref{fig:GS} are shown in Fig.~\ref{fig:GSplots}.
The colour conventions are as before: in particular the light green region is excluded by dijet pair searches.
The bound from $NN\to \pi \pi$ was computed as in Eq.~(\ref{eq:din-gen}) and is not as stringent as the present limit from $n-\bar{n}$, a feature that we observed in the previous models too. Since the model is MFV by construction, flavour violating processes are very well under control and we did not obtain
any relevant flavour constraints.
As a consequence -- besides the LHC limit on the squark masses $\gtrsim 400$ GeV -- the only relevant bound on the model is $n-\bar{n}$ itself, at least for $\lambda^{''}_{udb} \gtrsim 10^{-4}$, cf.~the right plot of the figure, and a large enough sbottom LR mixing.
As we can see, ESS has the potential of testing sbottom masses up to 2 TeV.
We did not show values of $\lambda^{''}_{udb}$ larger than 0.2, as searches for resonant single squark production at the LHC already exclude the model up to multi-TeV squarks for such large degree of RPV \cite{Monteux:2016gag}.

\subsection*{Model $\hbox{CK}$, spectrum $ \tilde \chi^\pm, \tilde b_R, \tilde t_R, \tilde b_L, ( \tilde t_L) $, couplings $\lambda^{''}_{tdb}, (A_b - \mu \tan\beta), (A_t - \mu \cot\beta) $}
\begin{figure}[t]
\begin{center}
\includegraphics[width=0.4\textwidth]{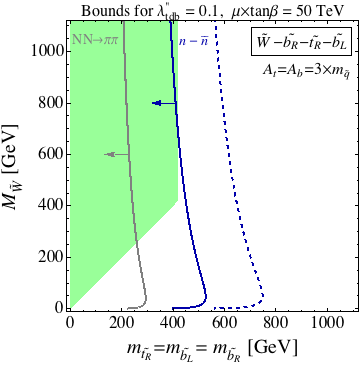}
\hspace{0.4cm}
\includegraphics[width=0.4\textwidth]{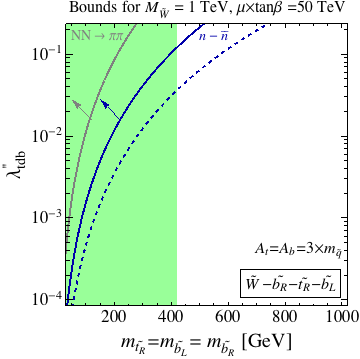}
\caption{Bounds and $n-\bar{n}$ prospects for
the Model $\hbox{CK}$. The colour code is as in Fig.~\ref{fig:ludb-d13}. \label{fig:CK}}
\end{center}
\end{figure}
In Fig.~\ref{fig:CK}, we show the results corresponding to the contribution of Fig.~\ref{fig:CandK}.
The colour code is as before. The novelty of this model with respect of the previous ones is that it involves $\lambda^{''}_{tdb}$. We checked that the analogous contribution with $\lambda^{''}_{tds}$ gives quantitatively similar results, with a slightly smaller numerical value of the oscillation probability.
The Chang and Keung contribution thus gives the very interesting possibility of testing through Baryon number violation different RPV couplings.
On the other hand, collider constraints are very similar to the previous case.

Another peculiar feature of the model is the dependence of $n-\bar{n}$ on the LR stop mixing (and thus on $A_t$) and on LH and RH stop masses.
Hence, direct links to the Higgs mass prediction and to considerations about fine tuning are therefore possible, though we omit such discussions in this work. 
However, for illustration purposes, we set the value of $A$-terms to be three times the squark mass. This choice maximizes the contribution to $n-\bar{n}$ without raising further constraints from possible charge- and colour-breaking minima of the scalar potential.
Despite this, we see that the induced $n-\bar{n}$ oscillation is numerically more suppressed than in the previous model and can be of phenomenological relevance only for sub-TeV squarks and large values of the RPV coupling, $\lambda^{''}_{tdb}=\mathcal{O}(0.1)$, not far from the present limits from resonant squark production \cite{Monteux:2016gag}.

\section{Non-renormalizable operators}
\begin{figure}[t]
\begin{center}
\includegraphics[width=.9\textwidth]{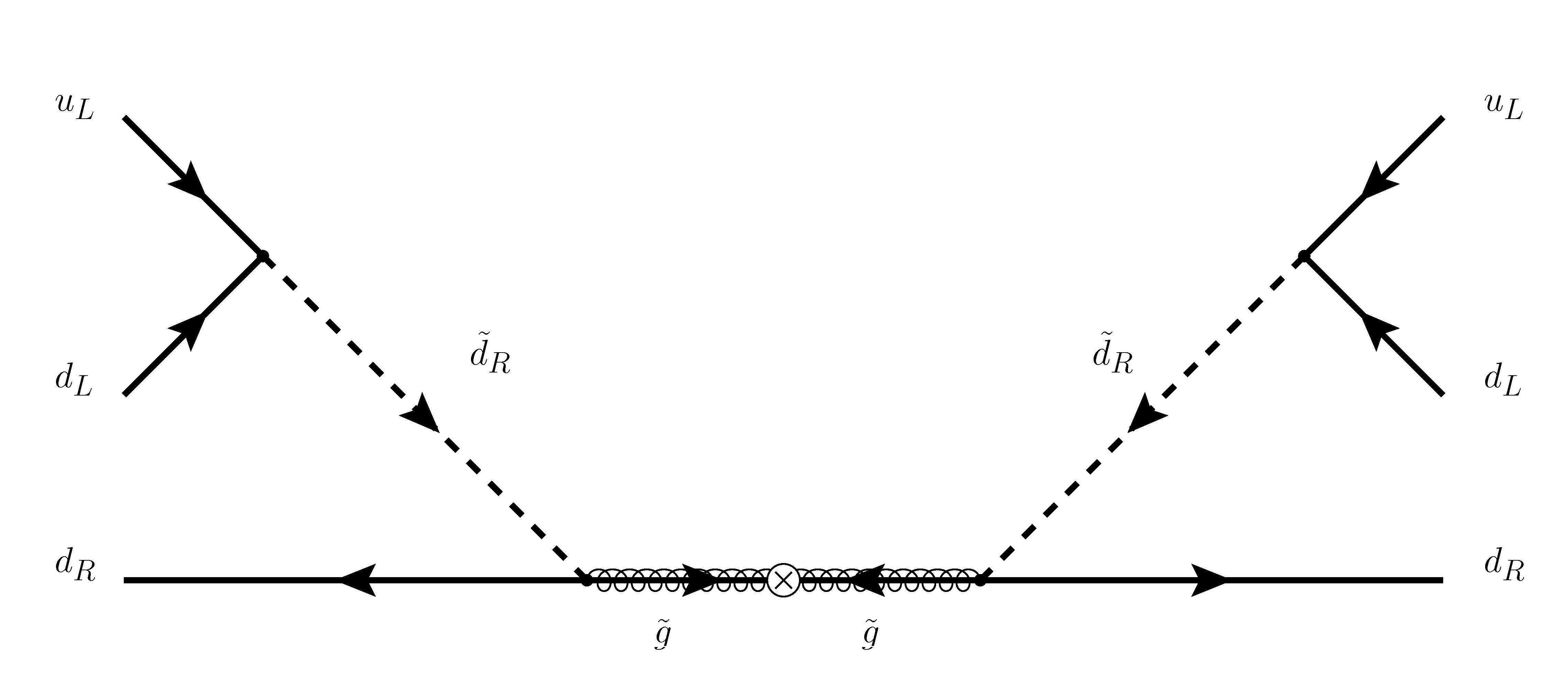}
\caption{The diagram contributing to $n-\bar{n}$ through the non-renormalizable operator pointed out in \cite{Csaki1}. \label{fig:CKV}}
\end{center}
\end{figure}
The body of this paper has concentrated on the effects of the renormalizable (dimension four) RPV operators. Given the smallness of their couplings, it is interesting to ask whether there might exist non-renormalizable operators that give a competing contribution. 
This possibility has been put forward by Csaki, Kuflik and Volansky in \cite{Csaki1}, see also \cite{Csaki2}. 
Considering the case of dimension five operators, one sees that there are two types of $|\Delta B| = 1$ operators that can be constructed out of MSSM superfields.

The first one is the cubic non-holomorphic (K\"ahler) term, 
\beq
           \frac{1}{2}\eta''_{ijk}[ \bar{D}^\dagger_i \; Q_j \cdot Q_k]   =\eta''_{ijk}[\bar{D}^\dagger_i \; U_{(j} D_{k)} ], 
 \label{hdKahler}
\eeq
where we have indicated the contraction of $SU(2)_L$ isospin indices with a dot and the usual antisymmetrization of the color indices with $[\dots]$.
In this case $\eta''_{ijk}$ must be symmetric in the last two family indices $j,k$ without any symmetry in the first index $i$. Thus the most relevant term for $n-\bar n$ oscillations is the one induced by $\eta''_{111}\equiv \eta''_{dud}$. %, giving rise to the operator ${\mathcal{O}}'_5$ of our classification (cf.~the Appendix). 
The oscillation amplitude  will now be suppressed not by a small flavour mixing but by the UV scale $M$ at which this operator is generated. 

The K\"ahler term in Eq.~\eqref{hdKahler} always gives rise to a squark-quark-quark coupling proportional to the ratio $\epsilon_d = m_d/M$, contributing to the six-quark operator ${\mathcal{O}}'_1$ of our classification (cf.~the Appendix). Furthermore, if one takes into account SUSY breaking by adding a spurion field $X$ to the operator in Eq.~\eqref{hdKahler}, with a non-vanishing $F$-term component $F_X$, one obtains an additional coupling, proportional to the ratio $\epsilon_X \equiv F_X/M^2$, contributing to the operator ${\mathcal{O}}'_5$. Thus we obtain the following six-quark operators
\beq
     {\mathcal{L}}_{n\bar n} = C^{\mathrm{CKV}}_{n\bar n, d} (u_L d_L d_L)^2 + C^{\mathrm{CKV}}_{n\bar n, X} (u_L d_L d_R)^2,
\label{nnbarkahler}
\eeq
with
\beq
    C^{\mathrm{CKV}}_{n\bar n, d/X}  = \frac{4}{3}\frac{g_s^2(\eta''_{dud})^2 \epsilon_{d/X}^2}{m^4_{\tilde d} m_{\tilde g}},
%\label{CKV}
\eeq
where one chooses $\epsilon_d$ or $\epsilon_X$ depending on the mechanism under consideration.
Comparing with~\eqref{eq:CZ}, under the reasonable assumption that the matrix elements of the 
the different six-quark operators 
%operators $(u_R d_R d_R)^2$ and $ (u_L d_L d_R)^2$ 
are of the same order, we see that $n-\bar n$ oscillation experiments constrain the product $\eta''_{dud}\epsilon_{d/X}$ in the same way 
%to be of the same order of magnitude 
as the combination $\lambda_{udk}^{''} (\delta^d_{RR})_{k1}$ in Eqs.~\eqref{eq:CZ} and \eqref{eq:CZnum}.
The LHC phenomenology and bounds will be similar to the previously considered models, featuring gluinos and squarks as well.

Notice that, if both operators in Eq.~\eqref{nnbarkahler} are generated at the same scale $M$ and  proportional to $\eta''_{dud}$, 
in the range of $M$ for which the first operator (involving $m_d/M$) gives rise to an observable $n-\bar n$ oscillation rate, the second operator (involving $F_X/M^2$) generically dominates. 

The second dimension five operator is instead a quartic holomorphic contribution to the superpotential,
\beq
       \rho_{ijk}[H_d\cdot Q_i\;  Q_j \cdot Q_k] = \sqrt{2} \rho_{ijk} v_d [D_i\; U_{(j} D_{k)}].
       \label{HQQQ}
\eeq
Now, $\rho_{ijk}$ must transform in the eight-dimensional representation of the family symmetry $SU(3)$ and, in particular, $\rho_{111}=0$. This means that terms of this type will always require squark family mixing to give rise to $n-\bar n$ oscillations. Thus, the operator in Eq.~\eqref{HQQQ}, which contributes to the six-quark operator ${\mathcal{O}}'_1$, is suppressed by a flavor mixing parameter as well as a factor $v_d/M$, although it could still be a non-negligible contribution to $n-\bar n$ oscillations given the smallness of the renormalizable couplings.  

\section{Conclusions}
Violation of baryon number $B$ is required to explain baryogenesis and plays an important role  in many theories of physics beyond the SM, motivating searches for $B$ violating processes. The second run of the LHC will continue to push the high energy frontier  searching for such theories. However, as is well-known in the context of flavour physics and CP violation, it is of utmost importance to also push the low energy frontier by means of precision experiments, since they can probe energy scales of the new physics that goes well beyond the reach the high energy colliders. In terms of baryon number violation, the recently proposed $n-\bar n$ oscillation experiment at ESS provides a great opportunity to make progress on this important issue.

In this paper we revisited this issue in the context of $B$ violating supersymmetry and examined several simplified models giving rise to $\Delta B=2$ processes and the constraints put  on them by flavour physics, di-nucleon decays and previous $n-\bar n$ oscillation experiments. We also recasted LHC searches and extracted the corresponding constraints on these simplified models.

We showed that, in terms of these simplified models and with the projected sensitivity of the proposed $n-\bar n$ oscillation experiment, this experiment will have a reach that in some cases goes beyond the reach of the LHC as well as the other experiments, as it can probe gluino and squark masses in the multi-TeV range. Hence, this is a striking example of the complementarity between the high and low energy frontiers.

Let us end by stressing the importance of an improved calculation of the  $n-\bar n$ matrix elements, since the current uncertainty severely limits the predictivity. The calculation done so far, based on the MIT bag model, is over thirty years old  \cite{Rao:1982gt,Rao:1983sd}. We would like to encourage the lattice gauge theory community to calculate these six-quark matrix elements, see Ref.~\cite{Buchoff:2012bm} for a discussion and some preliminary results.

\section*{Acknowledgments}
We thank our colleagues in the collaboration ``Neutron--Anti-Neutron Oscillations at ESS" for discussions. 
We are grateful to Chang Sub Shin for pointing out a mistake in the coefficient of the six-quark operators for the non-EW case (Z and BM).
D.\,M.~is supported by the Swedish Research Council. C.\,P.~is supported by the Swedish Research Council under the contract 637-2013-475, by IISN-Belgium (conventions 4.4511.06, 4.4505.86 and 4.4514.08) and by the ``Communaut\'e Fran\c{c}aise de Belgique" through the ARC program and by a ``Mandat d'Impulsion Scientifique" of the F.R.S.-FNRS.

\appendix
\section{$\Delta B = 2$ operators}
The relevant operators for $n-\bar{n}$ oscillation are $\Delta B = 2$, $\Delta I_3 = -1$ objects constructed out of six quark fields of the first family (``$uudddd$''). There are two different classes of operators one may consider, depending on whether one performs the classification before or after EWSB. The class of operators of the first type is obviously more restrictive since we must impose the full $SU(3)_c\times SU(2)_L\times U(1)_Y$ invariance and is the one that is relevant for the discussion of BSM physics taking place at energies higher than the EW scale. Operators of both kinds have been studied in the literature since the 80's. In particular, paper \cite{Caswell:1982qs} classified them eliminating all redundancies and computed their renormalization at leading order. This computation has been recently checked and pushed to next-to-leading order in \cite{Buchoff:2015qwa}. In this section we present their relevant results in a self-contained way and slightly extend the classification.

As far as the body of the paper is concerned, only a small number of operators arise in these models of baryonic RPV, as seen in (\ref{relevantops}). They are identified with the following operators to be introduced in this appendix:
\beq
     \urdrdr = {\mathcal{O}}_1,  \;  \; \urdrdl = {\mathcal{O}}_5, \;  \;  \uldldr = {\mathcal{O}}'_5, \;  \;   \urdrsr = {\mathcal{D}}_1.
\eeq
(The last operator is included as it contributes to di-nucleon decay.)

Let us begin with the larger class obtained imposing only $SU(3)_c\times U(1)_{\mathrm{e.m.}}$ invariance. We construct these operators out of the two components spinors $u_{L,\alpha}^a$ $d_{L,\alpha}^a$ $u_{R\dot\alpha}^a$ and $d_{R\dot\alpha}^a$ where, as usual, $\alpha, \dot\alpha$ are Weyl indices and $a=1,2,3$ is a colour index. Fierzing allows us to reduce all tensor structures to scalar fermionic bilinears and the only colour invariant combination will involve two $\epsilon$ tensors. We denote, for any of the fields above $[ \psi\; \psi'\; \psi''] \equiv  \epsilon_{abc}\psi^a \psi'^b \psi''^c$ and
${\acontraction{}{\psi}{}{\psi'}\psi\psi'} \equiv \epsilon_{\alpha\beta}\psi^\alpha\psi'^\beta~\mathrm{or}~\epsilon^{\dot\alpha\dot\beta}\psi_{\dot\alpha}\psi'_{\dot\beta}$. Notice that, due to the Grassmann
nature of the fields, ${\acontraction{[}{\psi}{}{\psi'}[\psi\psi'\dots]}=-{\acontraction{[}{\psi'}{}{\psi}[\psi'\psi\dots]}$, in particular ${\acontraction{[}{\psi}{}{\psi}[\psi\psi\dots]}=0$.

Let us begin, for illustration purpose and because they will be of interest in the discussion of specific models, with the operators that can be constructed out of only right-handed quarks. By inspection one can construct four non zero operators involving the set of fields $2\times u_R, 4\times d_R$:

\beqs
       A = \onecontr{u_R}{ d_R}{ d_R}{ d_R}{ u_R}{ d_R}&,~&
       B = \threecontr{u_R}{ d_R}{ d_R}{ d_R}{ u_R}{ d_R}\nn\\
       C =  \threecontr{u_R}{ d_R}{ d_R}{ d_R}{ d_R}{ u_R}&,~&
       D = \threecontrH{u_R}{ u_R}{ d_R}{ d_R}{ d_R}{ d_R}
\eeqs
These four operators however are \emph{not} linearly independent and e.g. $C$ and $D$ can be eliminated in favor of $A$ and $B$
\beq
     C = - A + B, \qquad\hbox{and}\qquad D = -2 A + B
\eeq
Notice that, because of the permutability of the fields inside $[\cdots]$ all Weyl contractions can be written either in the same form as operator $A$ or $B$.

Similar arguments can be repeated for any allowed combination of quarks.
The rules to construct these operators are the following. An allowed combination of six quark fields consists of two up-type quarks and four down-type quarks, of which an even number is left-handed (and thus also an even number is right-handed). For any allowed combination, first spilt the fields into all the inequivalent pairs of colour triples $[ \cdots ][ \cdots ]$ (after colour antisymmetrization, the Grassmann nature of the fields makes the order inside the $[\cdots]$ irrelevant). Then, take all possible Weyl  contractions $\acontraction{}{\cdot}{}{\;\cdot\;} \cdot\;\cdot\;$ between LH and RH quarks pairs eliminating those that yield zero by the considerations above. Lastly, find all linear dependences between the remaining operators.

At the end, we are left with 14 operators -- the following list of 7 operators $\mathcal{O}_i$ and those ($\mathcal{O}'_i$) constructed simply by exchanging $L$ and $R$ everywhere. The first two operators are just the operators $A$ and $B$ considered above.
\beqs
\mathcal{O}_1 = \onecontr{u_R}{ d_R}{ d_R}{ d_R}{ u_R}{ d_R} &,~&
\mathcal{O}_2 =  \threecontr{u_R}{ d_R}{ d_R}{ d_R}{ u_R}{ d_R} \nn\\
\mathcal{O}_3 = \onecontr{u_L}{ d_L}{ d_R}{ d_R}{ d_R}{ u_R} &,~&
\mathcal{O}_4 = \threecontr{u_R}{ d_R}{ d_L}{ u_L}{ d_R}{ d_R} \nn\\
\mathcal{O}_5 = \onecontr{u_R}{ d_R}{ d_L}{ d_L}{ u_R}{ d_R} &,~&
\mathcal{O}_6 = \threecontr{u_R}{ d_R}{ d_L}{ d_L}{ u_R}{ d_R} \nn\\
\mathcal{O}_7 = \threecontrH{d_R}{ d_R}{ u_L}{ u_L}{ d_R}{ d_R} &\phantom{,}~&
 \label{fulllist}
\eeqs
Note that all operators can be chosen such that the combination inside $[\dots]$ has the same valence as the neutron, in particular it is electrically neutral.

 We present, for comparison,  the conversion between the above basis and that used in~\cite{Rao:1982gt}:
\beqs
      \mathcal{O}_1 = \frac{1}{2}  \mathcal{O}3_{RRR} &, ~& \mathcal{O}_2 = \frac{3}{4}  \mathcal{O}3_{RRR} - \frac{1}{4}  \mathcal{O}2_{RRR} \nn\\
      \mathcal{O}_3 = -\frac{1}{2}  \mathcal{O}3_{RLR} &, ~& \mathcal{O}_4 = \frac{1}{4}  \mathcal{O}3_{RLR} - \frac{1}{4}  \mathcal{O}2_{RLR} \nn\\
      \mathcal{O}_5 = \frac{1}{2}  \mathcal{O}3_{RRL} &, ~& \mathcal{O}_6 = \frac{3}{4}  \mathcal{O}3_{RRL} - \frac{1}{4}  \mathcal{O}2_{RRL} \nn\\
      \mathcal{O}_7 = -\frac{1}{4}  \mathcal{O}1_{LRR} &\phantom{, }~&
\eeqs

We now want to analyze what kind of further restrictions are imposed by the requirement that these operators arise from operators invariant under the full EW group. For this we must combine the LH fields into a $SU(2)$ doublet $Q_L^i = (u_L, d_L)$ and introduce the Higgs field $H^i = (H^+, H^0)$. We also denote
$H_i \equiv \epsilon_{ij}H^j = (H^0, -H^+)$, $\tilde H^i = (-H^{0*}, H^{+*})$ and $\tilde H_i = (H^{+*}, H^{0*})$, so that after EWSB $H_i Q_L^i \to v u_L$ and $\tilde H_i Q_L^i \to v d_L $. Two $Q_L^i$ fields appearing in the same colour invariant ${[\dots]}$ and Weyl-contracted with each other can only be antysimmetrized with a $\epsilon_{ij}$, while the remaining must be contracted with the appropriate Higgs field and symmetrized.
It can easily be seen that no even-dimensional operator can be made neither with Higgs fields nor with covariant derivatives as it would break either Lorentz or weak isospin invariance. Since higher dimensional operators containing covariant derivatives are suppressed by additional powers of momenta, ($p/\Lambda_{\mathrm{BSM}}$) we will only look at those obtained with the Higgs field which are only suppressed by additional powers of $v/\Lambda_{\mathrm{BSM}}$.

All in all, we have four dimension 9 operators
\beqs
\mathcal{Q}_1 = \onecontr{u_R}{ d_R}{ d_R}{ d_R}{ u_R}{ d_R}  &,~&
\mathcal{Q}_2 = \threecontr{u_R}{ d_R}{ d_R}{ d_R}{ u_R}{ d_R} \nn\\
\mathcal{Q}_3 = \onecontr{Q_L^i}{ Q_L^j}{ d_R}{ d_R}{ d_R}{ u_R} \epsilon_{ij} &,~&
\mathcal{Q}_4 = \onecontr{Q_L^i}{ Q_L^j}{ d_R}{ d_R}{ Q_L^k}{ Q_L^l}\epsilon_{ij}\epsilon_{kl},
\label{dim9}
\eeqs
six dimension 11 operators
\beqs
\mathcal{Q}_5 = \onecontr{u_R}{ d_R}{ Q_L^i}{ Q_L^j}{ u_R}{ d_R} \tilde H_{(i} \tilde H_{j)} &,~&
\mathcal{Q}_6 = \threecontr{u_R}{ d_R}{ Q_L^i}{ Q_L^j}{ u_R}{ d_R} \tilde H_{(i} \tilde H_{j)} \nn\\
\mathcal{Q}_7 = \threecontr{u_R}{ d_R}{ Q_L^i}{ Q_L^j}{ d_R}{ d_R} H_{(i} \tilde H_{j)} &,~&
\mathcal{Q}_8 = \threecontrH{d_R}{ d_R}{ Q_L^i}{ Q_L^j}{ d_R}{ d_R} H_{(i} H_{j)} \\
\mathcal{Q}_9 = \onecontr{u_R}{ d_R}{ Q_L^i}{ Q_L^j}{ Q_L^k}{ Q_L^l} \tilde H_{(i} \tilde H_{j)} \epsilon_{kl} &,~&
\mathcal{Q}_{10} = \onecontr{ Q_L^i}{ Q_L^j}{ Q_L^k}{ Q_L^l}{ Q_L^m}{ Q_L^n} \tilde H_{(k} \tilde H_{l)}\epsilon_{ij}\epsilon_{mn}, \nn
\label{dim11}
\eeqs
where we denote by $(ij..)$ the symmetric combinations of $SU(2)_L$ indices.
Furthermore, we have three dimension 13 operators
\beqs
\mathcal{Q}_{11} &=& \threecontrH{ Q_L^i}{ Q_L^j}{ u_R}{ u_R}{Q_L^k}{Q_L^l} \tilde H_{(i} \tilde H_j \tilde H_k \tilde H_{l)} \nn\\
\mathcal{Q}_{12} &=& \threecontrH{ Q_L^i}{ Q_L^j}{ d_R}{ u_R}{Q_L^k}{Q_L^l} H_{(i} \tilde H_j \tilde H_k \tilde H_{l)} \nn\\
\mathcal{Q}_{13} &=& \threecontrH{ Q_L^i}{ Q_L^j}{ d_R}{ d_R}{Q_L^k}{Q_L^l} H_{(i} H_j \tilde H_k \tilde H_{l)}
\label{dim13}
\eeqs
and finally one dimension 15 operator
\beq
\mathcal{Q}_{14} = \threecontrH{ Q_L^i}{ Q_L^j}{Q_L^k}{Q_L^l}{Q_L^m}{Q_L^n} H_{(i} H_j \tilde H_k \tilde H_l \tilde H_m \tilde H_{n)}.
\eeq

After EWSB all the above $\mathcal{Q}$ operators reduce to linear combinations of the $\mathcal{O}$ operators multiplied by the appropriate powers of $v$:
\beqs
     &&  \mathcal{Q}_1 \to \mathcal{O}_1, \quad
       \mathcal{Q}_2 \to \mathcal{O}_2,\quad
       \mathcal{Q}_3 \to - 2 \mathcal{O}_3,\quad
       \mathcal{Q}_4 \to 4 \mathcal{O}'_5,\quad
       \mathcal{Q}_5 \to v^2 \mathcal{O}_5,\quad
       \mathcal{Q}_6 \to  v^2 \mathcal{O}_6, \nn\\
     &&       \mathcal{Q}_7 \to  v^2 \left( \frac{1}{2}\;\mathcal{O}_3 +  \mathcal{O}_4 \right),  \quad
       \mathcal{Q}_8 \to  v^2 \mathcal{O}_7,   \quad
       \mathcal{Q}_9 \to -2  v^2\mathcal{O}'_3, \quad
      \mathcal{Q}_{10} \to 4 v^2 \mathcal{O}'_1,   \nn\\
     &&  \mathcal{Q}_{11} \to v^4  \mathcal{O}'_7,   \quad
       \mathcal{Q}_{12} \to v^4 \left(\frac{1}{2} \; \mathcal{O}'_3 +  \mathcal{O}'_4\right),    \nn\\
      && \mathcal{Q}_{13} \to v^4 \left(-\; \mathcal{O}'_5 + \; \mathcal{O}'_6\right),  \quad
       \mathcal{Q}_{14} \to v^6 \left(-\frac{6}{5} \; \mathcal{O}'_1 +  \; \mathcal{O}'_2\right),
\eeqs

Operators involved in di-nucleon decay, with field content of type $uuddss$ can be classified along the same lines. We only present those constructed out of RH quarks. There are five independent ones that can be chosen as
\beqs
  && {\mathcal{D}}_1 = \onecontr{u_R}{d_R}{s_R}{s_R}{u_R}{d_R}, \quad  {\mathcal{D}}_2 = \onecontr{u_R}{s_R}{d_R}{s_R}{d_R}{u_R} \nn \\
  && {\mathcal{D}}_3 = \onecontr{u_R}{s_R}{d_R}{u_R}{s_R}{d_R}, \quad  {\mathcal{D}}_4 = \onecontr{s_R}{u_R}{u_R}{d_R}{d_R}{s_R} \nn \\
  && {\mathcal{D}}_5 = \threecontrH{s_R}{u_R}{u_R}{d_R}{d_R}{s_R}
\eeqs

When estimating these coefficients via lattice gauge theory one should renormalize from the BSM scale in which these operators are generated down to the nuclear scale. The renormalization coefficients of these operators have been computed in~\cite{Caswell:1982qs} and~\cite{Buchoff:2015qwa} and we use it to compute the suppression/enhancement in our basis. Denoting by $\mathcal{O}_i$ and  $\mathcal{O}_i^0$ the renormalized and bare operators respectively, we have, to LO in $\alpha_s$,
\beqs
   \begin{pmatrix} \mathcal{O}_1 \\  \mathcal{O}_2 \end{pmatrix} &=&  \begin{pmatrix} \mathcal{O}_1^0 \\  \mathcal{O}_2^0 \end{pmatrix}+
   \frac{\alpha_s}{\pi \epsilon} \begin{pmatrix}1 & 0\\-6 & 6 \end{pmatrix}  \begin{pmatrix} \mathcal{O}_1^0 \\  \mathcal{O}_2^0\end{pmatrix}\nn\\
     \begin{pmatrix} \mathcal{O}_3 \\  \mathcal{O}_4 \end{pmatrix} &=& \begin{pmatrix} \mathcal{O}_3^0 \\  \mathcal{O}_4^0 \end{pmatrix}+
   \frac{\alpha_s}{\pi \epsilon} \begin{pmatrix}-1 & 0\\2 & 3 \end{pmatrix}  \begin{pmatrix} \mathcal{O}_3^0 \\  \mathcal{O}_4^0\end{pmatrix}\nn\\
     \begin{pmatrix} \mathcal{O}_5 \\  \mathcal{O}_6 \end{pmatrix} &=& \begin{pmatrix} \mathcal{O}_5^0 \\  \mathcal{O}_6^0 \end{pmatrix}+
   \frac{\alpha_s}{\pi \epsilon} \begin{pmatrix}0& 0\\-3 & 3 \end{pmatrix}  \begin{pmatrix} \mathcal{O}_5^0 \\  \mathcal{O}_6^0\end{pmatrix}\nn\\
   \mathcal{O}_7 &=& \left(1 + 3 \frac{\alpha_s}{\pi \epsilon} \right) \mathcal{O}_7^0
\eeqs
It is clear that, because of chirality, the operators mix only in pairs, with the last one unmixed. The same is true for the primed operators $\mathcal{O}'_i$.

Defining
\beq
      \xi\! =\!  \left(\! \frac{\log \Lambda_{\mathrm{BSM}}/ \Lambda_{\mathrm{QCD}}}{\log m_t/ \Lambda_{\mathrm{QCD}}}\!  \right)^{1/14}
            \!   \left(\! \frac{\log m_t / \Lambda_{\mathrm{QCD}}}{\log m_b/ \Lambda_{\mathrm{QCD}}}\!  \right)^{3/46}
             \!  \left(\! \frac{\log m_b / \Lambda_{\mathrm{QCD}}}{\log m_c/ \Lambda_{\mathrm{QCD}}}\!  \right)^{3/50}
            \!  \left(\! \frac{\log m_c / \Lambda_{\mathrm{QCD}}}{\log M_N/ \Lambda_{\mathrm{QCD}}}\!  \right)^{1/18}
\eeq
describing the RG evolution from $\Lambda_{\mathrm{BSM}}$ to the nucleon mass $M_N$, (the exponents being equal to $1/(2(11 - 2/3 N_f))$ for the appropriate number of flavours), we get, choosing $ \Lambda_{\mathrm{BSM}} = 10\;\mathrm{TeV}$ and  $ \Lambda_{\mathrm{QCD}} = 200\;\mathrm{MeV}$
\beqs
         {\mathcal{O}_1}_{ |\Lambda_{\mathrm{BSM}}} &=& \xi^{-4} {\mathcal{O}_1}_{ |M_N}  =
                0.61\; {\mathcal{O}_1}_{ |M_N} \nn\\
        {\mathcal{O}_2}_{ |\Lambda_{\mathrm{BSM}}} &=& \xi^{-24} {\mathcal{O}_2}_{ |M_N} +
         \frac{6}{5}(\xi^{-4} - \xi^{-24}){\mathcal{O}_1}_{ |M_N} = 0.049\; {\mathcal{O}_2}_{ |M_N} +
         0.67\;{\mathcal{O}_1}_{ |M_N}  \nn\\
        {\mathcal{O}_3}_{ |\Lambda_{\mathrm{BSM}}} &=& \xi^{4} {\mathcal{O}_3}_{ |M_N}  =
                1.65\; {\mathcal{O}_3}_{ |M_N} \nn\\
        {\mathcal{O}_4}_{ |\Lambda_{\mathrm{BSM}}} &=& \xi^{-12} {\mathcal{O}_4}_{ |M_N} -
         \frac{1}{2}(\xi^{4} - \xi^{-12}){\mathcal{O}_3}_{ |M_N} = 0.22\; {\mathcal{O}_4}_{ |M_N} -
         0.71\; {\mathcal{O}_3}_{ |M_N}  \nn\\
        {\mathcal{O}_5}_{ |\Lambda_{\mathrm{BSM}}} &=& {\mathcal{O}_5}_{ |M_N} \nn\\
        {\mathcal{O}_6}_{ |\Lambda_{\mathrm{BSM}}} &=& \xi^{-12} {\mathcal{O}_6}_{ |M_N} +
        (1 - \xi^{-12}){\mathcal{O}_5}_{ |M_N} = 0.22 \;{\mathcal{O}_6}_{ |M_N} +
         0.68\;{\mathcal{O}_5}_{ |M_N}  \nn\\
        {\mathcal{O}_7}_{ |\Lambda_{\mathrm{BSM}}} &=& \xi^{-12} {\mathcal{O}_7}_{ |M_N}  =
                0.22\; {\mathcal{O}_7}_{ |M_N}
\eeqs
We see that the operator ${\mathcal{O}_3}$ is enhanced by $ 65\%$, ${\mathcal{O}_5}$ is unrenormalized to LO and the remaining operators are suppressed.

\bibliographystyle{mine}
\bibliography{bibliography}

\end{document}